# The amazing dynamics of stochastic pattern formation and growth models inspired by the Conway's Game of Life

Leonid P. Yaroslavsky,
Dept. of Physical Electronics, School of Electrical Engineering,
Tel Aviv University, Tel Aviv 69978, Israel,
yaro@eng.tau.ac.il

## *Abstract*

Several modifications of the famous mathematical "Game of Life" are introduced by making "Game of Life" rules stochastic and mutual influence of cells in their 8-neighborhood on a rectangular lattice spatially non-uniform. Results are reported of experimental investigation of evolutionary dynamics of the introduced models. A number of new phenomena in the evolutionary dynamics of the models and collective behavior of patterns they generate are revealed, described and illustrated: formation of maze-like patterns as fixed points of the models, "self-controlled" growth, "eternal life in a bounded space" and "coherent shrinkage".

## 1. Introduction.

In this paper we consider evolutional dynamics of pattern formation and growth models derived through modifications of the famous mathematical model known as Conway's "Game of Life" [1]. In this model, 2D arrays of binary, i.e. assuming values 1( "live") or 2 ("empty"), cells arranged in nodes of a rectangular lattice within a rectangular "vital space" of a finite size (see

Figure 1) are subjected to evolution.

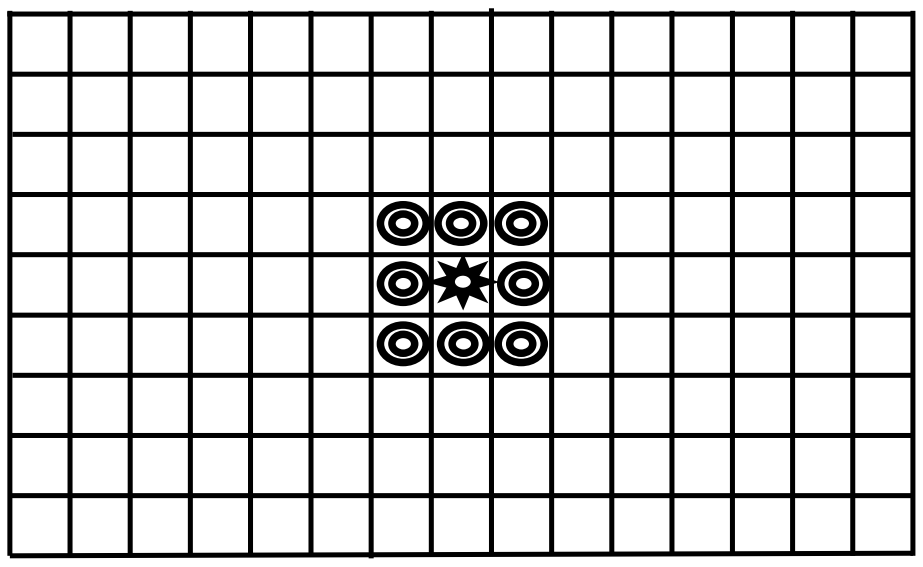

Figure 1. Cells in a "vital space" on a rectangular lattice. 8-neigboring cells of a cell indicated by the 8-point star are indicated by donuts

Each cell in this array has 8 neighbors (

Figure 1). Rules of the evolution are as following. At each subsequent step of evolution,

(i) each "empty" cell that has exactly 3 "alive" neighbor cells in its 3x3 neighborhood on the rectangular lattice gives a "birth", i.e. becomes "live";

(ii) each "live" cell that has less than 2 and more than 3 "alive" cells in the neighborhood dies, i.e. becomes "empty";

(iii) in all other cases nothing happens.

Patterns generated by this model in course of evolutional steps $t, (t = 1,2,....)$ can formally be described by the equation:

$$pattern^{(t)}(k,l) = \{pattern^{(t-1)}(k,l)\delta[S_8^{(t-1)}(k,l) - 2]\} \forall \{\delta[S_8^{(t-1)}(k,l) - 3]\};$$
$$pattern^{(0)}(k,l) = seed_pattern(k,l) \quad (1)$$

where $(k,l)$ are integer indices of cells on the rectangular lattice, $\delta(\cdot)$ is the Kronecker delta ($\delta(0) = 1; \delta(x \neq 0) = 0$), $S_8^{(t-1)}(k,l)$ is the sum of values of 8 neighboring cells of (***k,l***)-th cell on the lattice, $seed_pattern(k,l)$ is an arbitrary initial binary pattern used as a seed, and $\{.\}\forall\{.\}$ denotes element-wise logical "OR" operation on arrays of binary numbers.

The remarkable property of the “Game of life” is that, from an arbitrary seed patters, it produces, in course of evolution, three types of pattern:

- stable patterns, which, once appeared, remain unchanged unless they collide with neighbor patterns, which can happen in course of evolution;
- periodical patterns (“oscillators”), which repeat themselves after a certain number of the evolution steps; obviously, stable patterns can be regarded as a special case of periodical patterns with a period of one step.
- self-replicating moving patterns (“gilders”), which move across the lattice and replicate themselves in a shifted position after a certain number of steps; this can be regarded as a general “space-time” periodicity.

Since its invention, “Game of life” has been intensively studied experimentally by numerous enthusiasts, which have been competing between each other in discovery of new stable patterns and oscillators. As a result, very many types of stable patterns, “oscillators” and “gliders” have been discovered, including very sophisticated ones such as “Gosper Glider Gun” and “2C5 Space Ship Gun P690”. These patterns as well as their classification issues and rates of appearance one can find elsewhere ([2-9]).

“Game of life” is not only a splendid plaything for mathematicians and amateurs. It can also, after appropriate modifications, serve as a base for evolutionary 2D pattern formation and growth models, and, more generally, for models of 2D nonlinear dynamic systems with feedback. We pursued this option in Refs. 10-13. Specifically, we

- interpreted “Game of life” in terms of nonlinear dynamic systems with feedback, and showed that it can algorithmically be regarded as akin to pseudo-random number generators and to earlier stochastic growth models by M. Eden ([14-16])
- introduced a stochastic modification of the Conway’s model, in which death of “live” cell with less than 2 and more than 3 neighbors occurs with a certain probability $\boldsymbol{P_{death} \leq 1}$ , the case of $\boldsymbol{P_{death} = 1}$ being correspondent to the standard non-stochastic Conway’s model,
- demonstrated that modified in this way model tends to produce, in course of evolution, maze-like patterns with chaotic dislocations, which very much remind zebra skin and tiger fur patterns, fingerprints, magnetic domain patterns and alike; 1D version of the model produces patterns that remind those some see-shells develop in their life.

- Introduced a “gray-scale” modification of the model, in which cells can assume arbitrary values between zero and one and logical Conway’s Game of Life rules are replaced by fuzzy logics.

In this paper, we introduce several new modifications of the standard Conway’s model and describe results of computer simulation experiments, which reveal new phenomena in the evolutionary dynamics of the models.

As seed patterns, several realizations of arrays of pseudo-random numbers with different rate of “live” cells as well as several “solid” seed patterns were used in the experiments. They are shown in Figure 2.

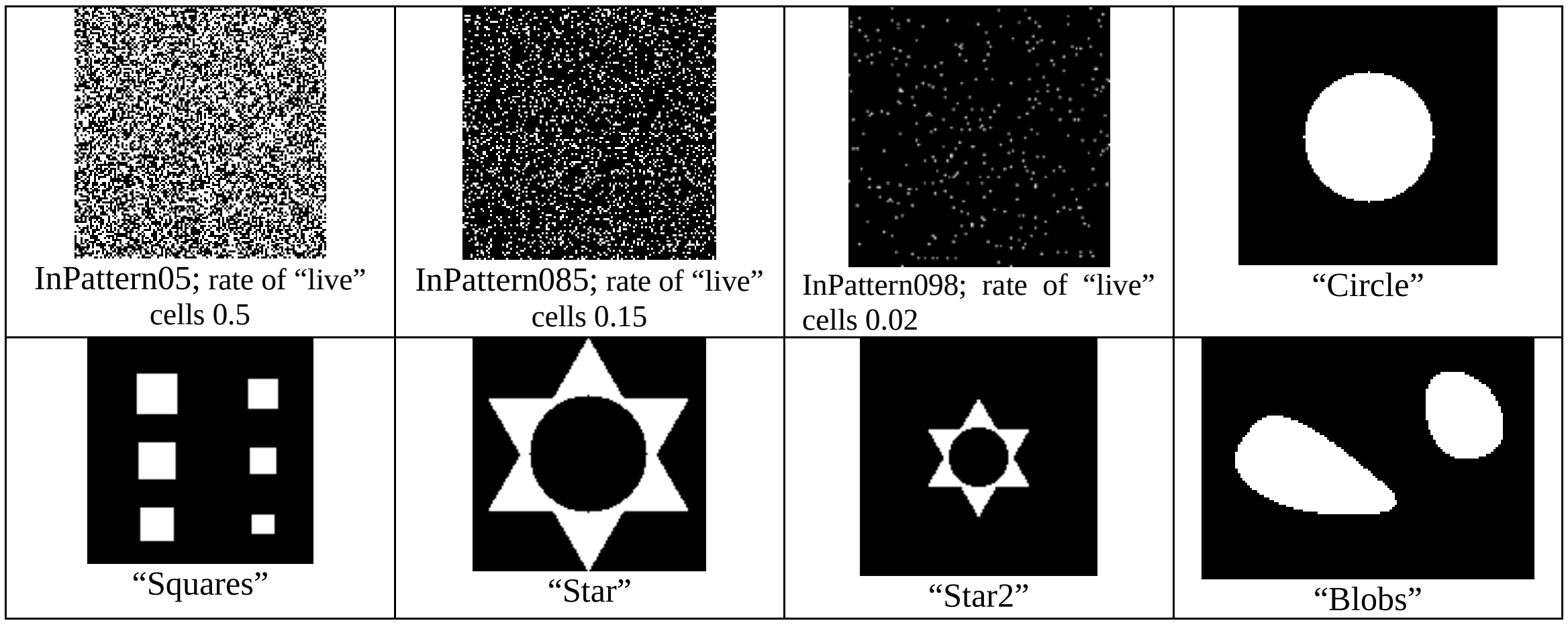

Figure 2. Examples of seed patterns used in the experiments. “Live” cells are shown white, empty cells are shown black)

## 2. Conway’s Game of Life modifications

Considered in this paper modifications of the standard Conway’s model given by Eq. 1 are made in two ways:

(i) “Deaths” of “live” cells that have less than two and more than three “live” neighbors are made stochastic with a probability $P_{death} \leq 1$.

(ii) Counting the number of “live” cells in 8-neighborhood of each cell by means of summation $S_8^{(t-1)}(k,l)$ of their binary values is replaced by a weighted summation with rounding up the summation result:

$$\tilde{S}_w^{(t-1)}(k,l) = Round\left(\sum_{m=-1}^{1}\sum_{n=-1}^{1} w_{m,n}\, pattern^{(t-1)}(k-m,l-n)\right), \qquad (2)$$

where weights $\{w_{m,n}\}$ are entries of a 3x3 weight matrix $\mathbf{W}$

$$\mathbf{W} = \{w_{m,n}\} = \begin{bmatrix} w_{-1,-1} & w_{-1,0} & w_{-1,1} \\ w_{0,-1} & 0 & w_{0,1} \\ w_{1,-1} & w_{1,0} & w_{1,1} \end{bmatrix}, \qquad (3)$$

which defines the model. Weight matrices selected for our study are presented in Table 1. For convenience of further referencing, we call them "masks".

**Table 1**

| | | | | |
|---|---|---|---|---|
| Standard Conway's_mask | Isotropic_mask | Diagonal_mask | Cross_mask | Cross4_mask |
| $\begin{bmatrix} 1 & 1 & 1 \\ 1 & 0 & 1 \\ 1 & 1 & 1 \end{bmatrix}$ | $\begin{bmatrix} 0.7 & 1 & 0.7 \\ 1 & 0 & 1 \\ 0.7 & 1 & 0.7 \end{bmatrix}$ | $\begin{bmatrix} 1 & 0.7 & 1 \\ 0.7 & 0 & 0.7 \\ 1 & 0.7 & 1 \end{bmatrix}$ | $\begin{bmatrix} 0.3 & 1 & 0.3 \\ 1 & 0 & 1 \\ 0.3 & 1 & 0.3 \end{bmatrix}$ | $\begin{bmatrix} 0 & 1 & 0 \\ 1 & 0 & 1 \\ 0 & 1 & 0 \end{bmatrix}$ |
| Cross4diag_mask | Hex0_mask | Hex1_mask | Hex2_mask | |
| $\begin{bmatrix} 1 & 0 & 1 \\ 0 & 0 & 0 \\ 1 & 0 & 1 \end{bmatrix}$ | $\begin{bmatrix} 1 & 0 & 1 \\ 1 & 0 & 1 \\ 1 & 0 & 1 \end{bmatrix}$ | $\begin{bmatrix} 0.75 & 0.5 & 0.75 \\ 1 & 0 & 1 \\ 0.75 & 0.5 & 0.75 \end{bmatrix}$ | $\begin{bmatrix} 1 & 0.75 & 0.5 \\ 0.75 & 0 & 0.75 \\ 0.5 & 0.75 & 1 \end{bmatrix}$ | |

"Isotropic" mask is introduced with a purpose of securing better than in the standard model correspondence of mutual influence of the cells to Euclidian distance between them in a real 2D space. "Diagonal mask" is a $45^o$ rotated "isotropic" mask. In the "Cross" mask, influence of more distant diagonal cells is further decreased comparing to the "Isotropic" mask, and in the "Cross4" mask it is completely eliminated. "Cross4diagonal" mask is a $45^o$ rotated version of the "Cross4" mask.

Masks "Hex0" and "Hex1" are introduced in an attempt to simulate cell neighborhoods that consists of 6 cells instead of the standard 8 cell neighborhood. Mask "Hex2" is a $45^o$ rotated version of the mask "Hex1".

## 3. *Ordering of chaos: the standard model*

We start with the more or less known type of the evolutionary dynamics, the "ordering of chaos". By the "ordering of chaos" we mean formation, out of, generally, chaotic seed patterns, stable formations that are "fixed points" of the model, i.e. formations, which, once appeared, are either not changing or oscillating in space or in "time" in course of evolution . In this section we revisit the "ordering of chaos" phenomenon for the standard model.

Obviously, fixed points of the model must be patterns that consist of cells with only two or three "live" neighbors, which would not die on the next step of evolution, and of "empty" cells with more than or less than three "live" neighbor cells, which would not come to life. Stable and oscillating patterns generated by the standard non –stochastic model, i.e. for the probability of "death" $\boldsymbol{P_{death}=1}$ , are well known and well reported. However, it turns out that such patterns are rather the exception than the rule because they appear only when $\boldsymbol{P_{death}}$ is strictly equal to one. Experiments show that, as soon as $P_{death}$ becomes only a little less than one, oscillating formations characteristic for the non-stochastic standard Conway's model occasionally collapse, producing chaotic clouds of "live" and "dying" cells that collide with each other and do not seem stabilizing ever. As $\boldsymbol{P_{death}}$ goes down further, these clouds are becoming denser in the "vital space", gradually fill in it keeping their "births/deaths" activity seemingly permanently and demonstrating kind of "eternal life".

For the probability of "death" lower than $\boldsymbol{P_{death}\approx 0.5}$, substantial changes in the evolutionary dynamics of the model are becoming noticeable. After a certain number of evolutionary steps, in different parts of the "vital space" patches of stripy patterns of different size and orientation emerge and grow in the see of active chaos. Borders of the patches remain to be active until $\boldsymbol{P_{death}}$ is higher than about 0.3. When $\boldsymbol{P_{death}}$ becomes lower than about 0.3, patches borders tend to stabilize, after a certain number of the evolutionary steps, the higher the lower $\boldsymbol{P_{death}}$, into mature maze-like patterns consisting of patches of alternative stripes of "live" and "empty" cells chaotically interrupted by dislocations, in which the direction of stripes is either switched to the perpendicular one or stripes of "live" and "empty" cells switch their positions. An example of such patterns is shown in **Figure 3**, right image. These patterns are, obviously, fixed points of the standard non-stochastic model as well. Therefore, for the stochastic modification of the standard Conway's model "ordering of "chaos" is possible only if the probability of "death" is sufficiently small ($P_{death} <\approx 0.3$). For higher probabilities of "death" the models exhibits the "eternal life" dynamics.

| Standard Conway's mask model; Seed pattern InPattern05 | **$P_{death} \equiv 1$** Isolated individual stable patterns and "oscillators" | **$0 < P_{death} <\approx 0.3$** Maze-like formations with "chaotic" dislocations |
|---|---|---|
| | 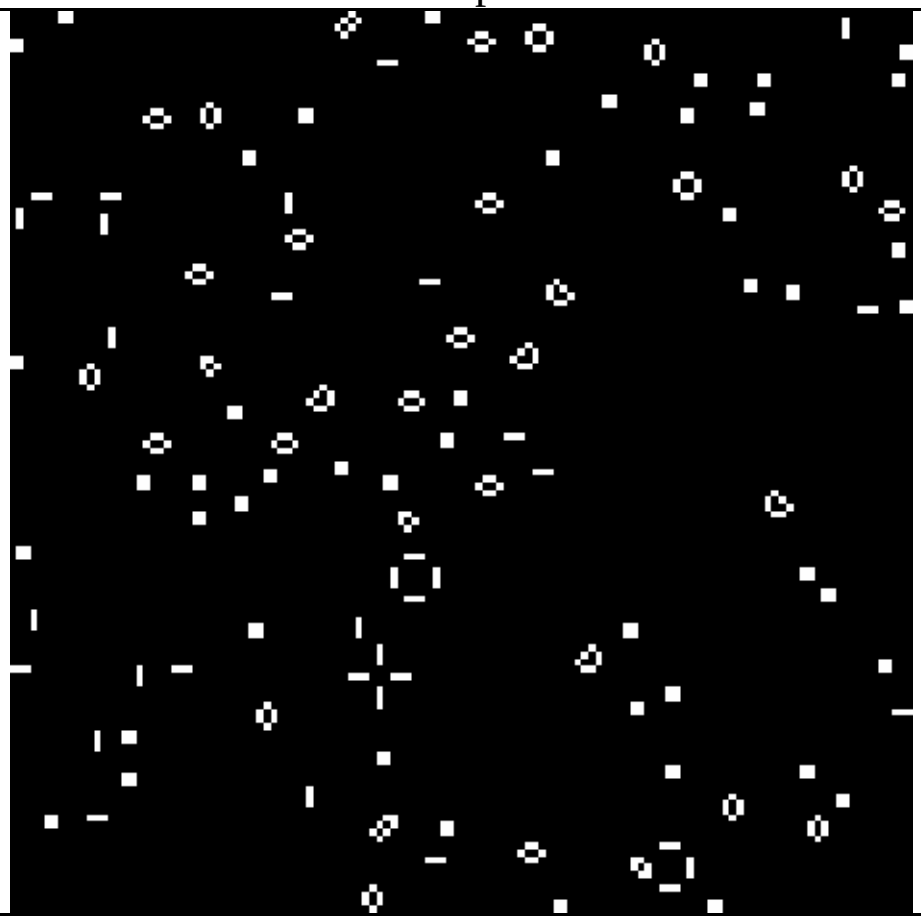 | 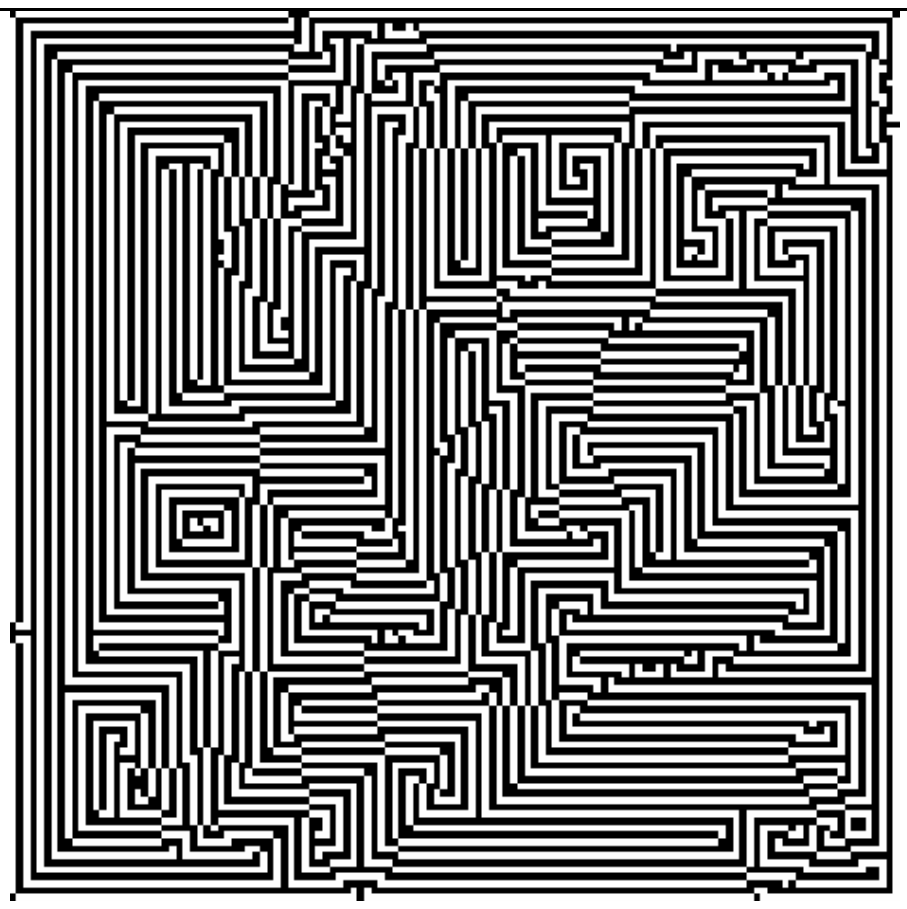 |

Figure 3. "Ordering of chaos": formation of individual stable patterns and maze-like stripy patterns. "Live" cells are shown white and "empty" cells are shown black.

| | | | |
|---|---|---|---|
| Standard Conway's mask model; Seed pattern InPattern098 | 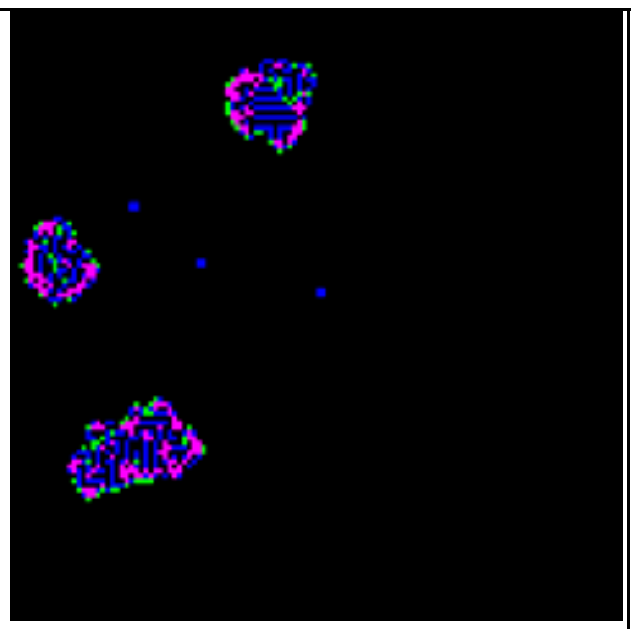 $P_{death}$=0.25; 25 evsteps. | 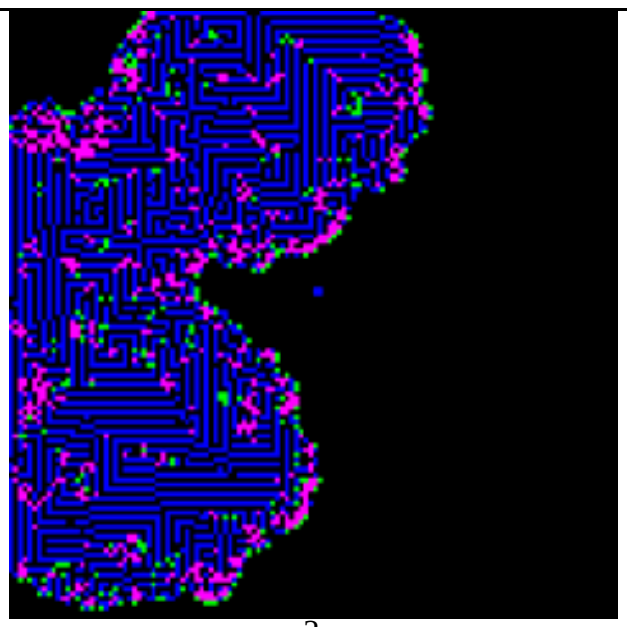 $P_{death}$=0.25; $10^2$ evsteps. | 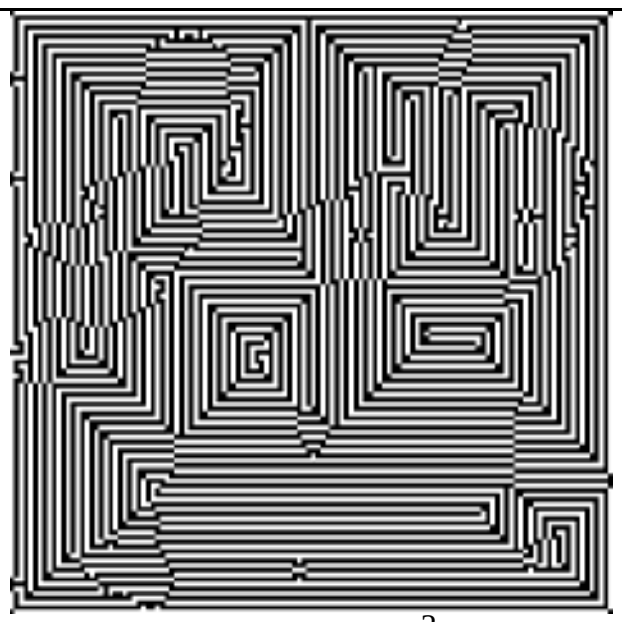 $P_{death}$=0.25; $5.9 \times 10^3$ evsteps. Stable |
| Standard Conway's mask model; Seed pattern "Star" | 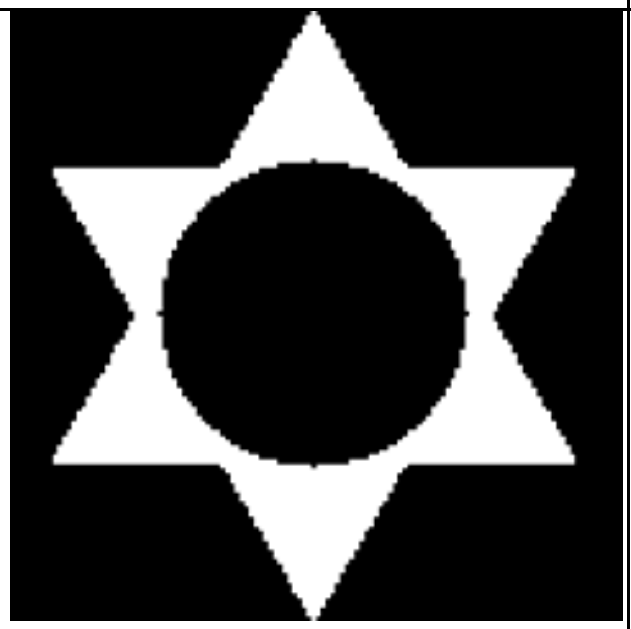 | 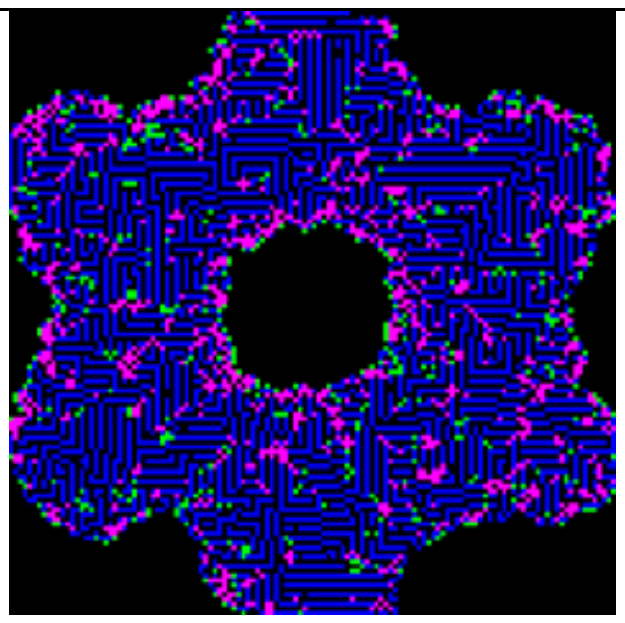 $P_{death}$=0.25; 40 evsteps | 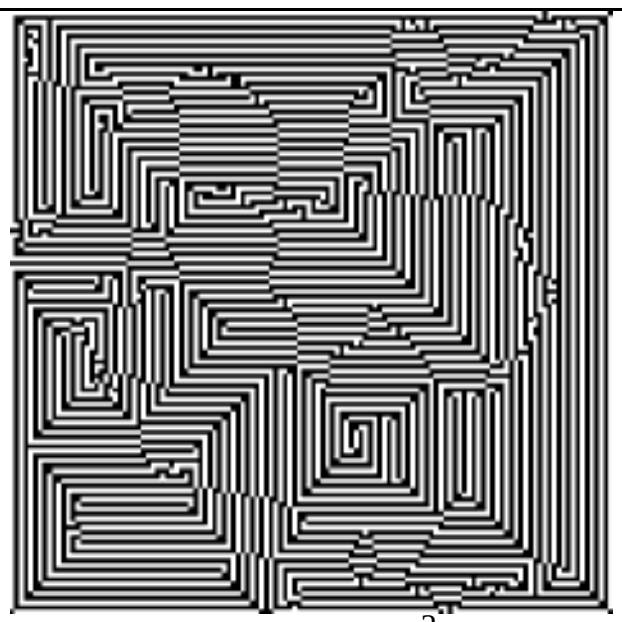 $P_{death}$=0.25; $4.5 \times 10^3$ evsteps. Stable |

Figure 4. Growth of maze-like patterns from sparse or solid seed patterns**.** In color coded images cells that will "die" on the next step are shown pink, cells, in which "birth" will take place on the next step, are shown green, "live" stable cells are shown blue and empty cells are shown black.

Genesis of emerging and growing of the maze-like patters can be better seen when seed patterns are sparse or solid. For sparse pseudo-random seed patterns, few isolated seeds of growth occasionally emerge that start growing and merging each other and gradually fill in the entire vital space. For solid seed patterns, growth starts at the pattern borders and then quite rapidly, in terms of the number of evolutionary steps, propagates to empty parts of the vital space. These processes are illustrated in

**Figure 4.**

Emergence of maze-like patterns as fixed points of the standard Conways' model for low probability of "death" has been already reported earlier ([10-13]). In the experiments reported in this paper, a new remarkable property of the maze-like stable patterns generated by the stochastic modification of the standard Conway's model was observed, their capability to grow and merge. If one takes, as a seed pattern, a fragment of a maze-like stable pattern or a maze-like stable pattern with a hole and let them evolve with the probability of "death" $\boldsymbol{P_{death} <\approx 0.25}$, the former will grow until it fills in the entire vital space and the latter will grow to fill the hole, as it is illustrated in

Figure 5. One can also implant fragments of one maze-like stable pattern into another and use the pattern with the implanted fragment as a seed pattern for further evolution of the model. After some number of evolutionary steps, implanted fragment perfectly tailors itself in the new "home" as it is shown in Figure 6.

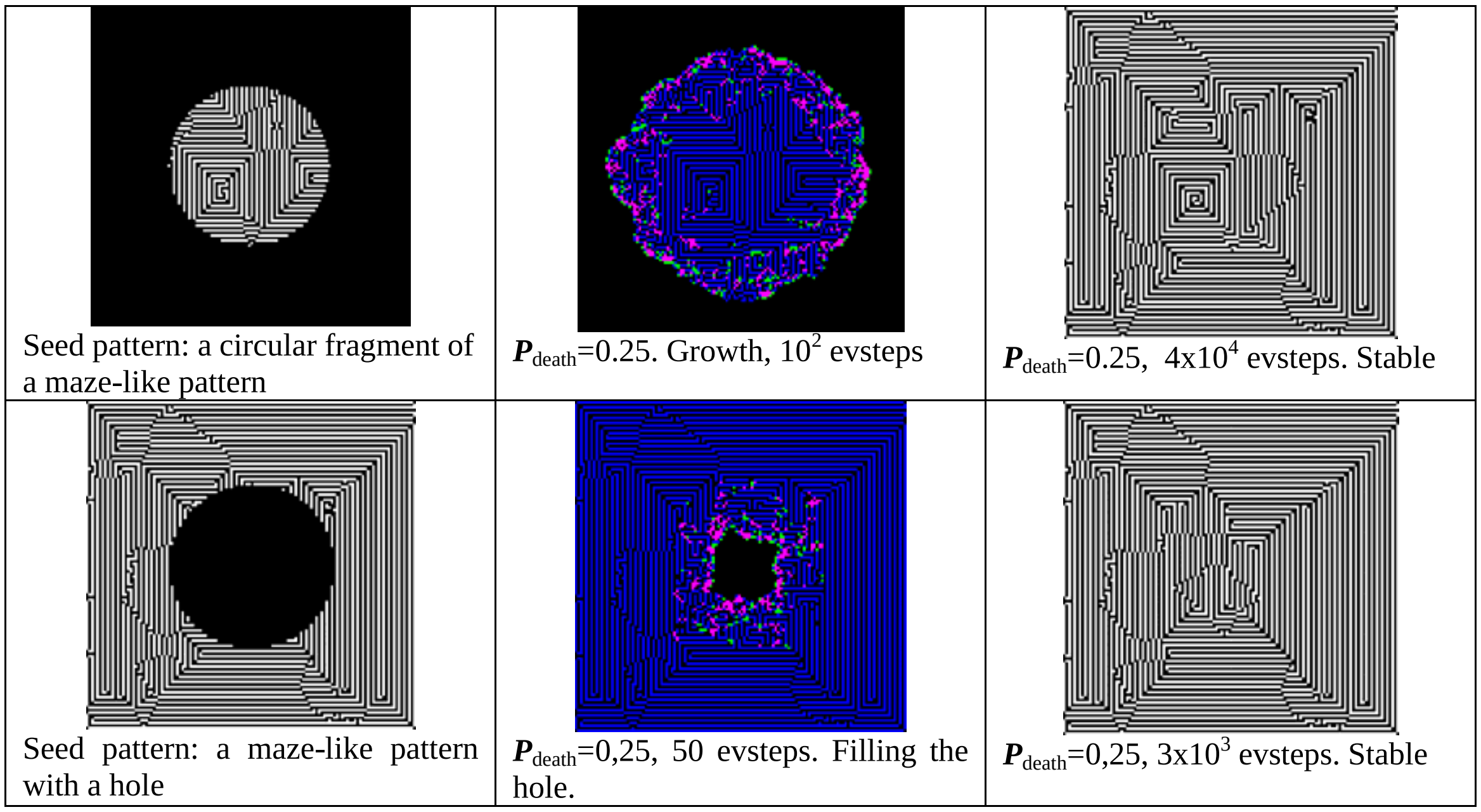

Figure 5. Growth capability of the maze-like patterns generated by the standard model. Upper row: growth of a circular fragment of a stable stripy pattern; bottom row: filling a circular hole in a fragment of a stable maze-like pattern. In color coded images, cells that will "die" on the next step are shown red, cells to give birth are shown green, "live" cells are shown blue and empty cells are shown black. In black and white images, "live" cells are shown white.

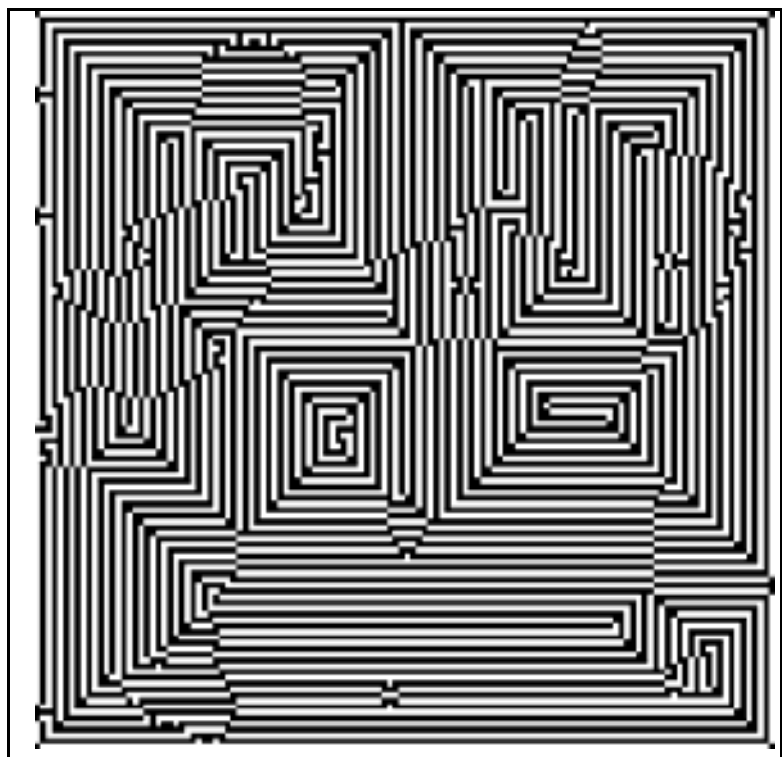

Maze-like pattern ***SP1***.

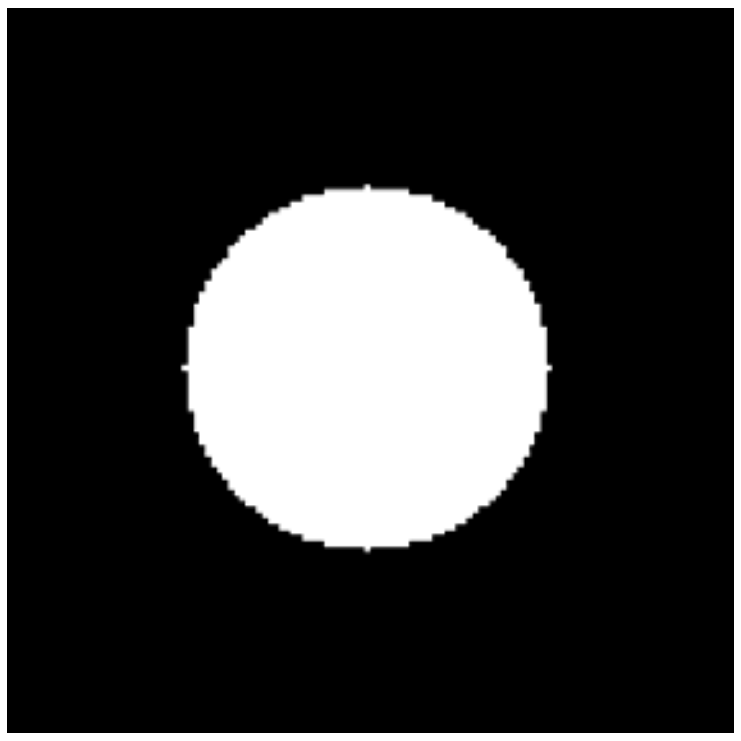

Circular window ***CW*** (while – 1, black-0) -

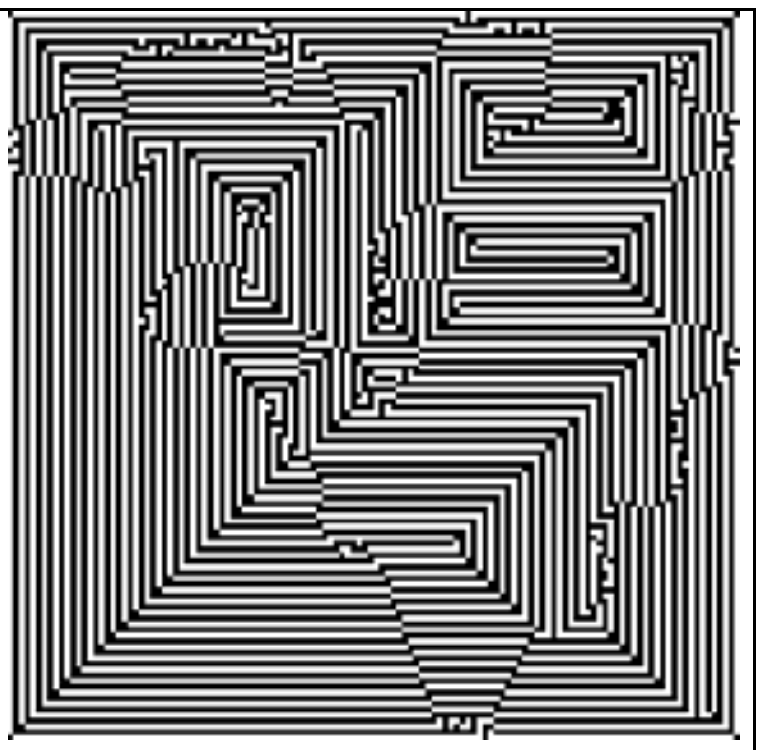

Maze-like pattern ***SP2***.

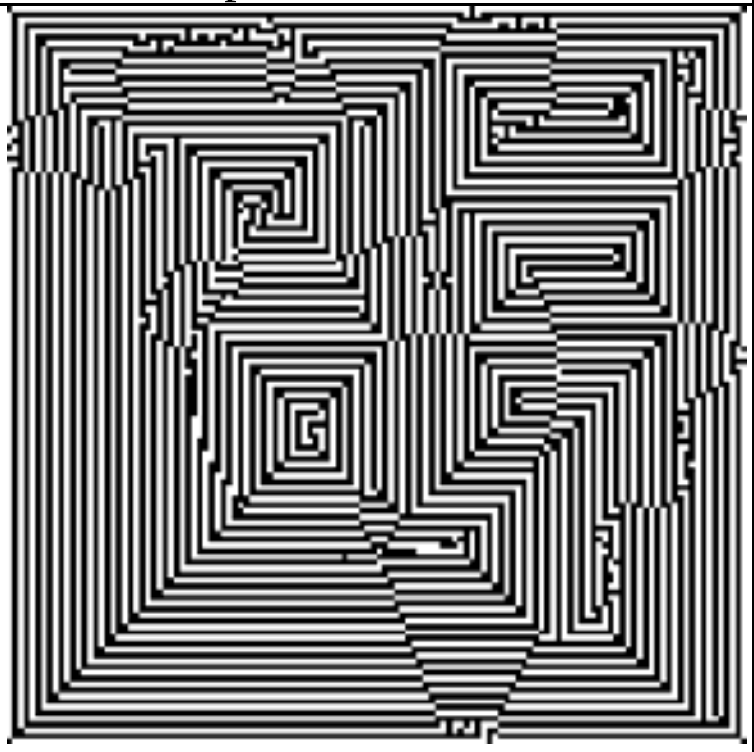

Seed pattern ***SP1.*CW+SP2_2.*(1-CW)***

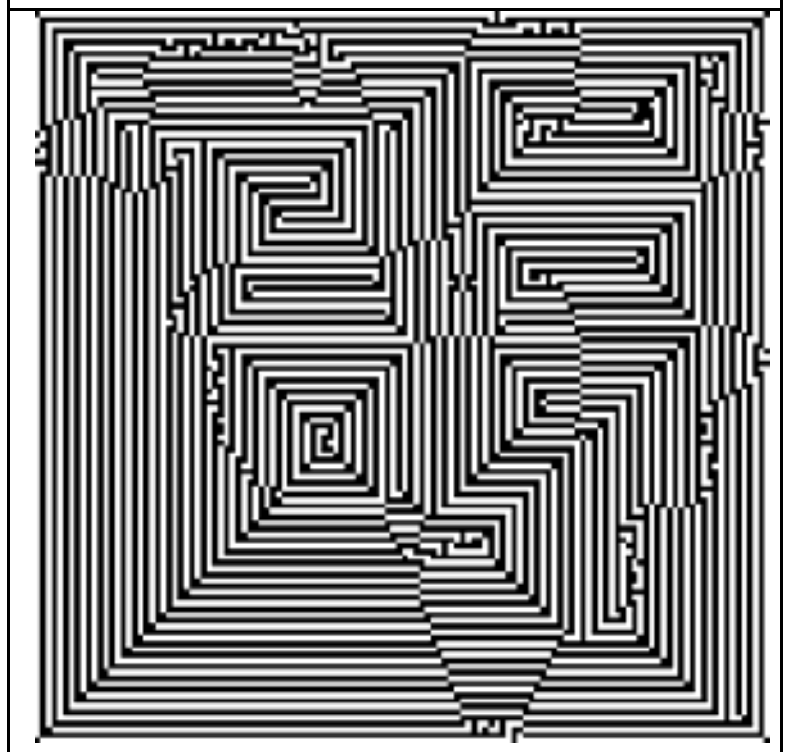

Stable pattern evolved from the seed pattern ***SP1.*CW+SP2.*(1-CW)*** .

Figure 6. Implanting a fragment of one maze-like pattern into another. Initial patterns are shown in the upper row. Second pattern with implanted central circular fragment of the first pattern and the resulting from it as a seed pattern the new stable maze-like pattern are shown in the bottom row. Circular mask used for extracting the implanted pattern is shown in the middle.

## 4. "Ordering of chaos": other models

For $P_{death} = 1$, the other models introduced in Sect. 2 demonstrate "ordering of chaos" dynamics similar to that of the standard stochastic Conway's model: their evolution ends up with a set of isolated stable formations illustrated in Figure 7 or oscillating patterns. Examples of oscillating patterns observed in the experiments are shown in Figure **8** (for Isotropic, Diag, Cross and Cross4 masks) and in Figure 9 (for Hex0, Hex1 and Hex2 masks).

| **Standard mask** | **Isotropic_mask** | **Diagonal_mask** |
|---|---|---|
| 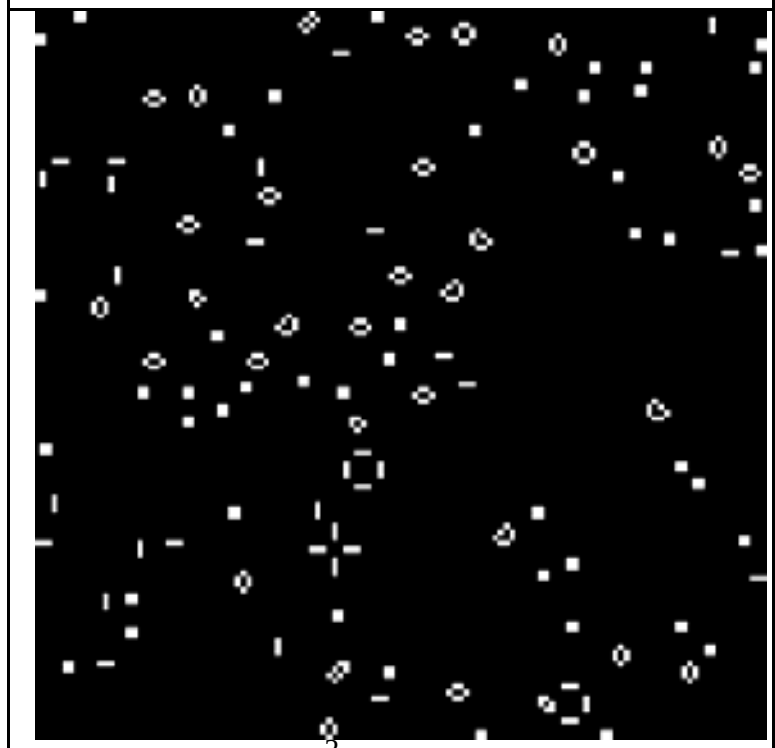<br>$\boldsymbol{P}_{death}$=1; 1.7x$10^3$ evsteps;<br>Stable with oscillators | 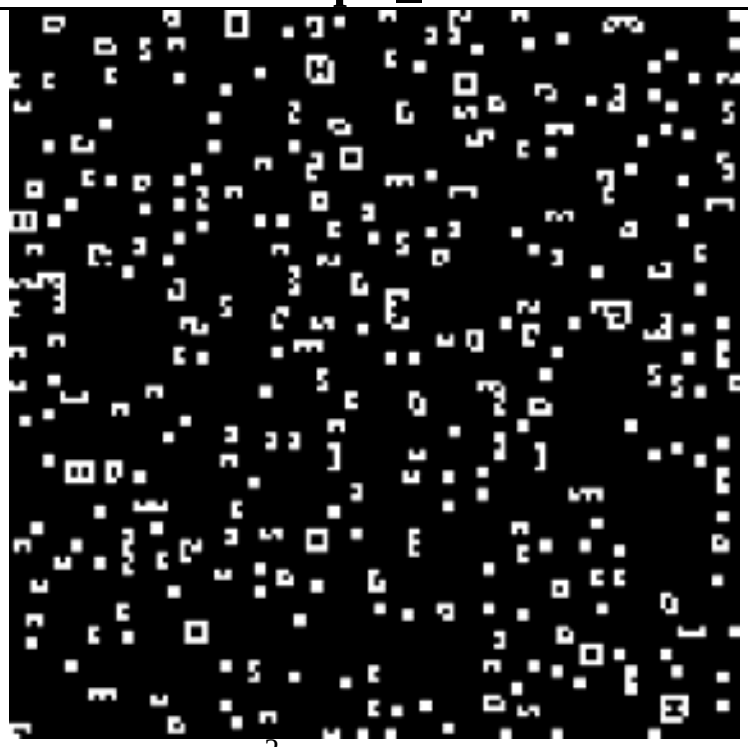<br>$\boldsymbol{P}_{death}$=1; 2x$10^2$ evsteps;<br>Stable with oscillators | 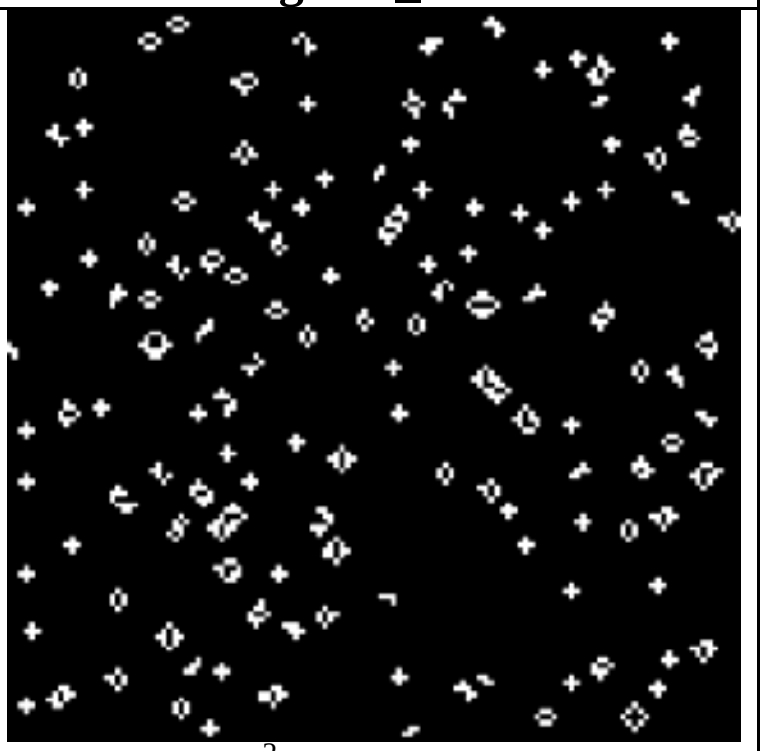<br>$\boldsymbol{P}_{death}$=1; 2x$10^2$ evsteps;<br>Stable with oscillators |
| **Cross_mask** | **Cross4_mask** | **MaskCross4diag** |
| 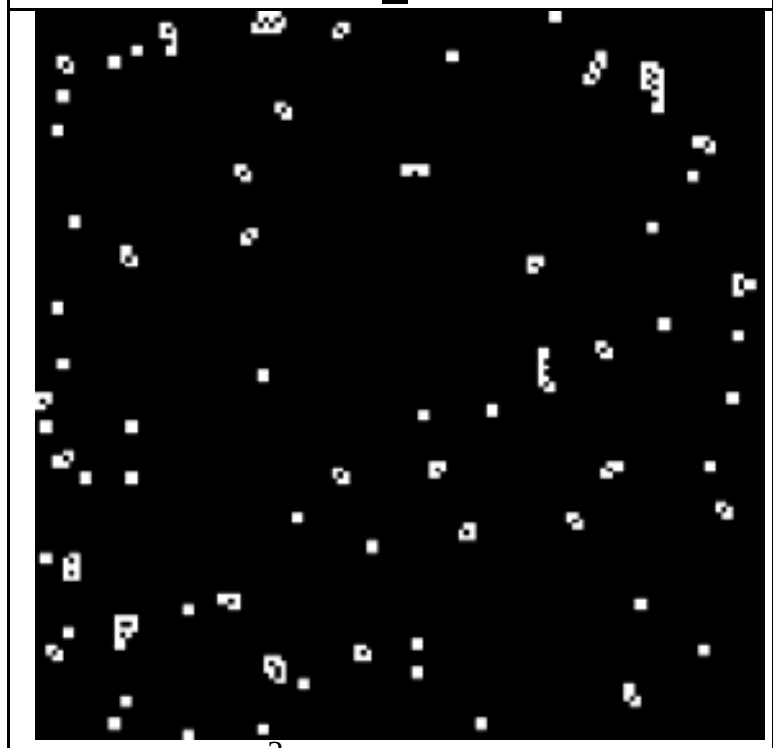<br>$\boldsymbol{P}_{death}$=1; $10^2$ evsteps;<br>Stable with oscillators | 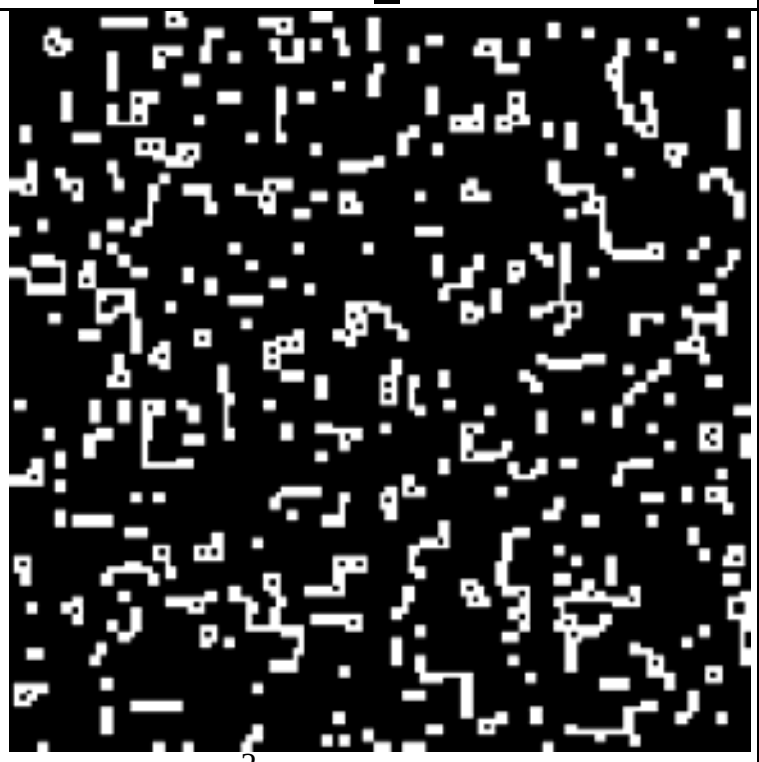<br>$\boldsymbol{P}_{death}$=1; $10^2$ evsteps.<br>Stable with 3 oscillators | 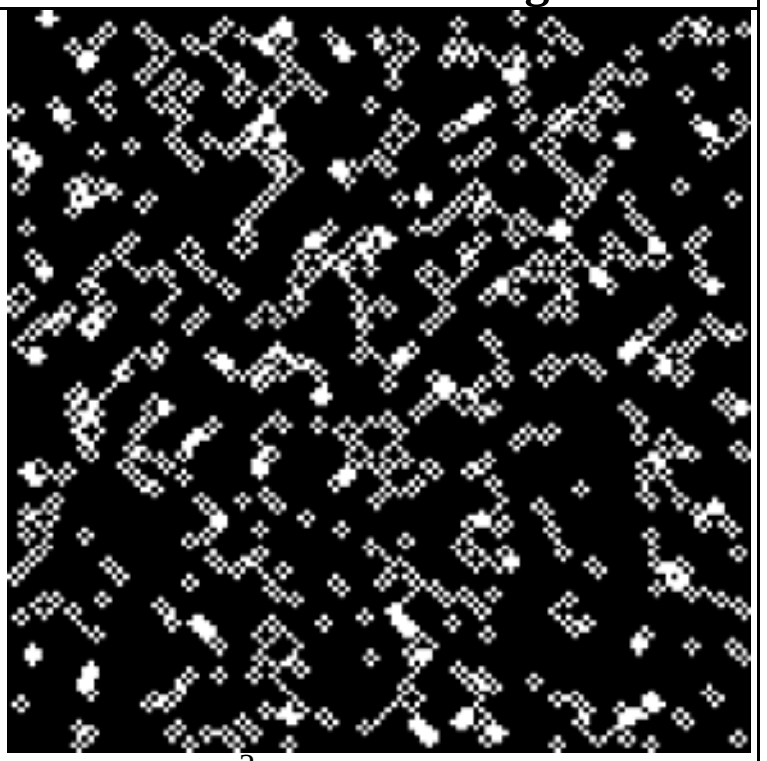<br>$\boldsymbol{P}_{death}$=1; $10^2$ evsteps;<br>Stable with 1 oscillator |
| **Hex0-mask** | **Hex1-mask** | **Hex2-mask** |
| 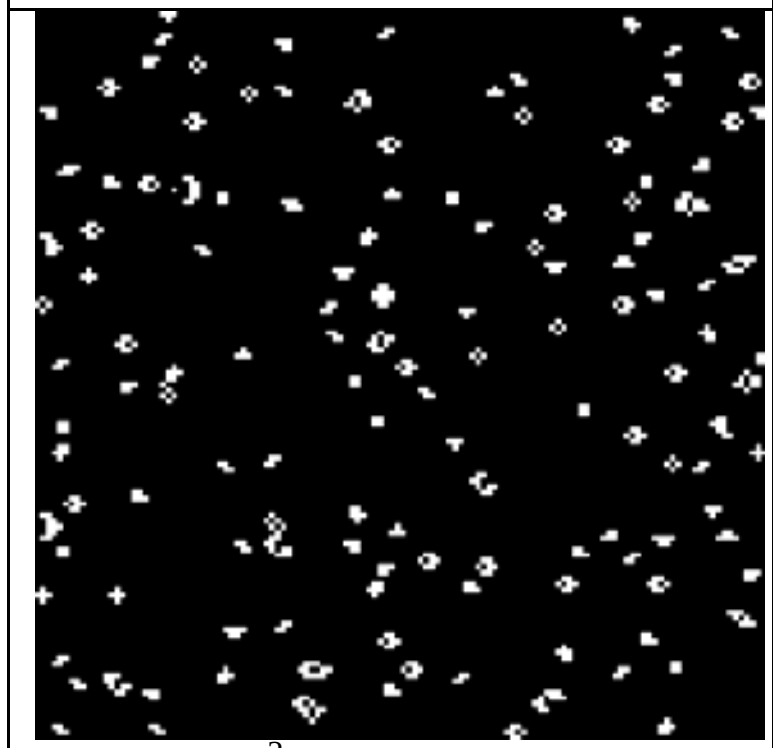<br>$\boldsymbol{P}_{death}$=1; $10^2$ evsteps;<br>Stable with several oscillating formations | 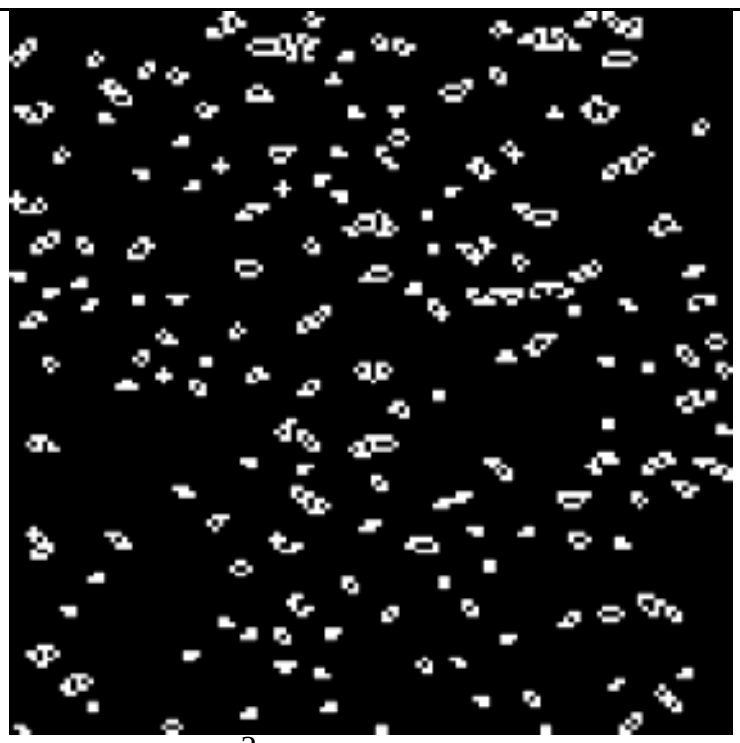<br>$\boldsymbol{P}_{death}$=1; $10^2$ evsteps;<br>Stable | 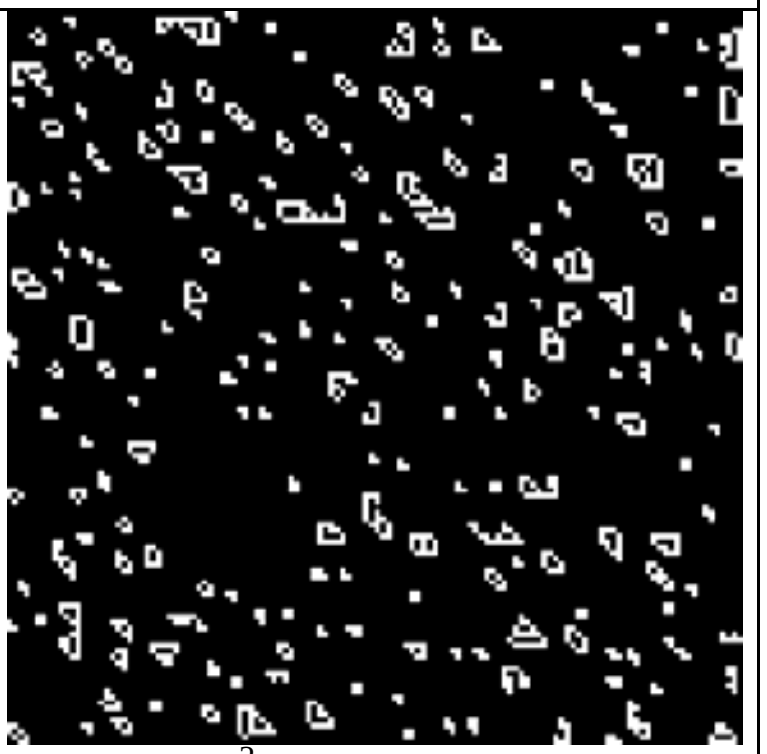<br>$\boldsymbol{P}_{death}$=1; $10^2$ evsteps;<br>Stable with oscillators |

Figure 7. Stable points of the considered models for $\boldsymbol{P_{death} = 1}$ (pseudo random seed pattern “InPattern05”). “Live” cells are shown white, empty cells are shown black.

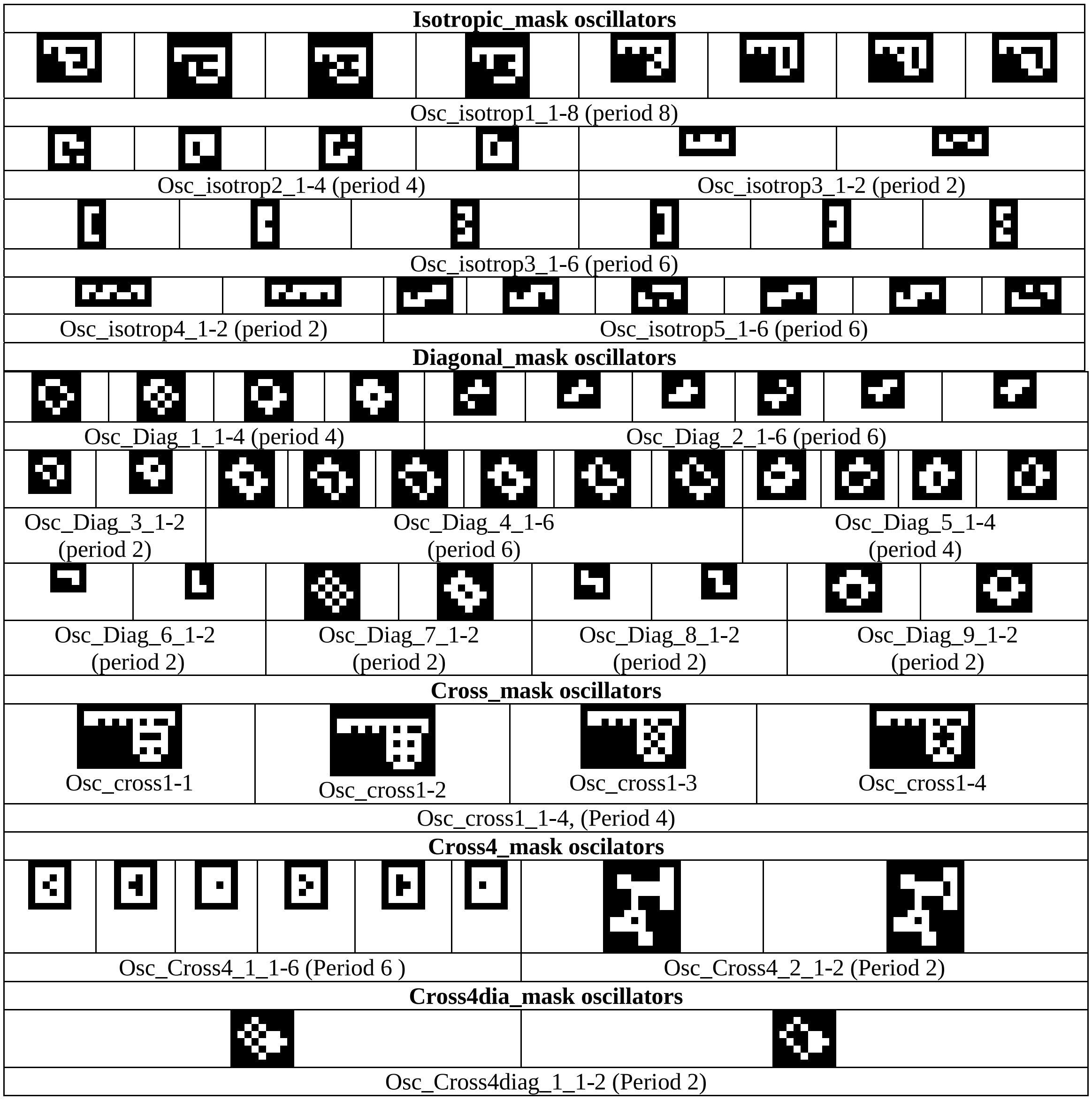

Figure 8. Samples of oscillating patterns and their phases for Isotropic, Diagonal, Cross, Cross4 and Cross4diag masks. “Live” cells are shown white, empty cells are shown black.

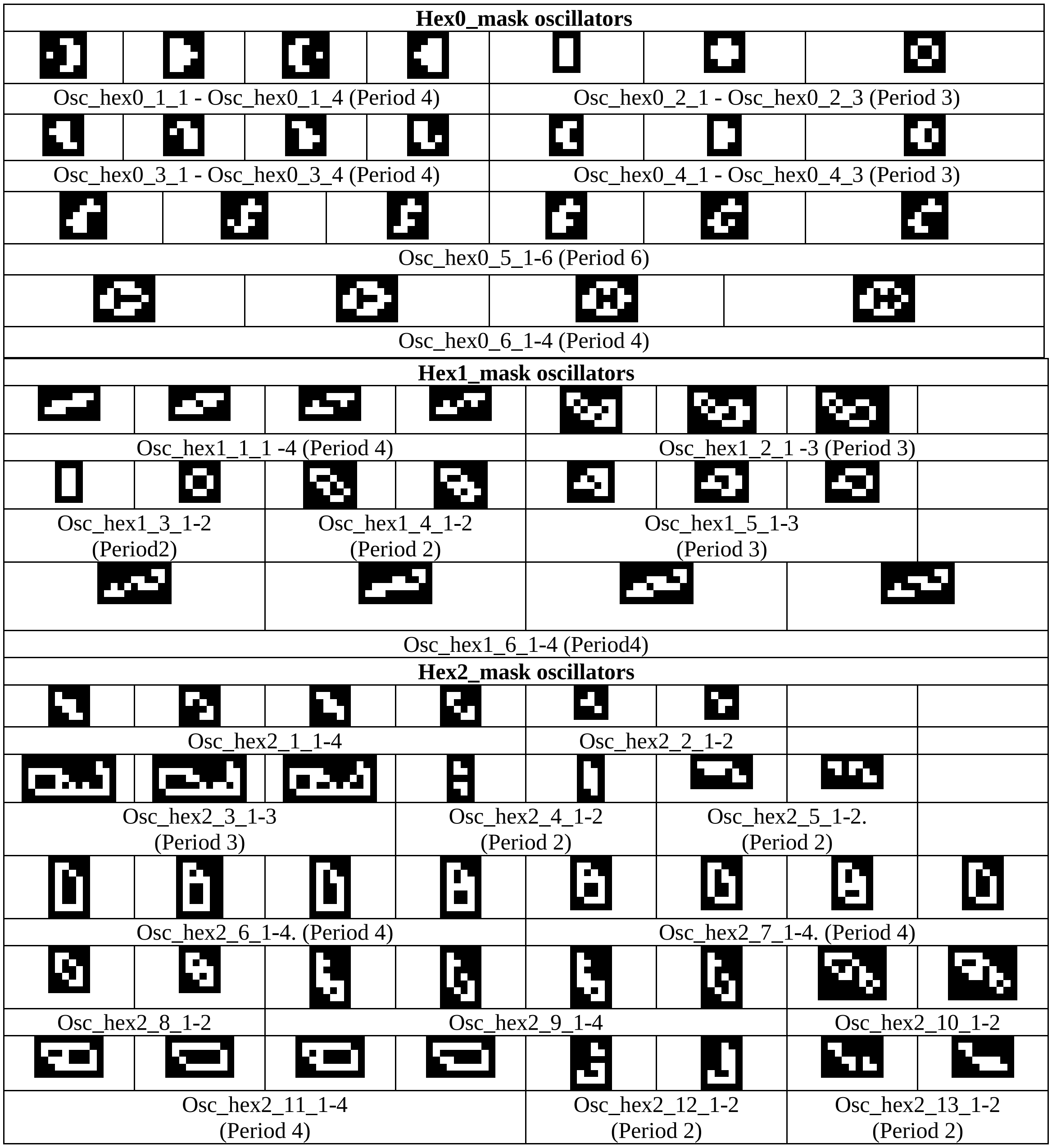

Figure 9. Samples of oscillating patterns and their phases for Hex0, Hex1 and Hex2 masks. “Live” cells are shown white, empty cells are shown black.

For $P_{death} < 1$, “ordering of chaos” type of the evolutionary dynamics similar to that for the standard model was observed only for models with “Isotropic”_mask and Hex2_mask: for $\approx 0.85 < P_{death} <= 1$, they end up with isolated individual stable formations and for $0 < P_{death} <\approx 0.5$, they converge to stable maze-like patterns (see Figure 10).

| **Isotropic_mask** | |
|---|---|
| $\approx 0.85 < P_{death} <= 1$<br>Isolated individual stable patterns | $0 < P_{death} <\approx 0.5$<br>Maze-like formations with “random” dislocations |
| 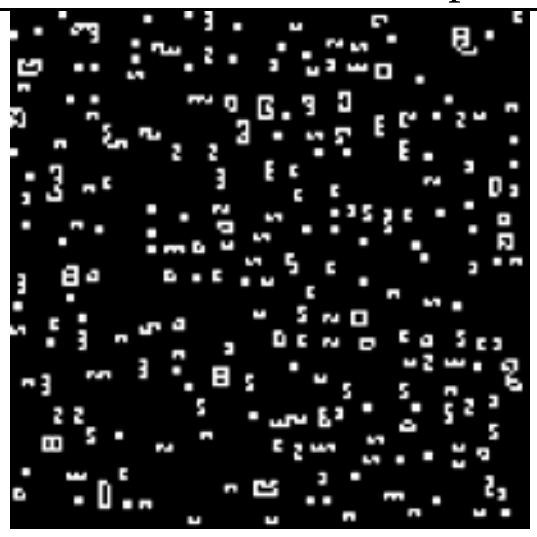<br>Seed pattern “InPattern05”. $P_{death}$=0.85; $2.1\text{x}10^2$ evsteps. Stable | 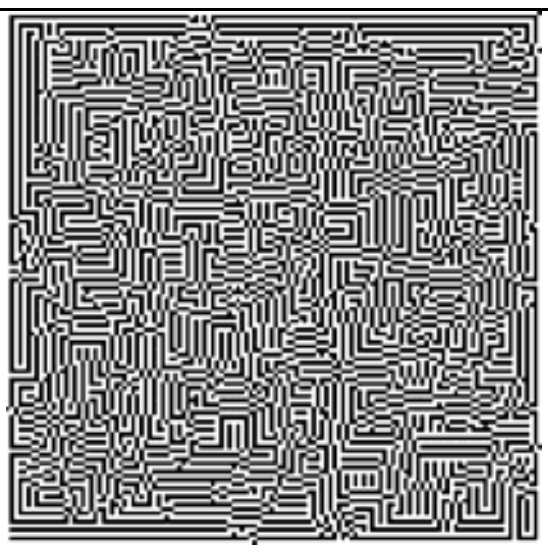<br>Seed pattern “InPattern05”. $P_{death}$=0.4; $2\text{x}10^2$ evsteps; Stable |
| **Hex2_mask** | |
| $\approx 0.85 < P_{death} <= 1$<br>Isolated individual stable patterns | $0 < P_{death} <\approx 0.5$;<br>Maze-like formations with “random” dislocation |
| 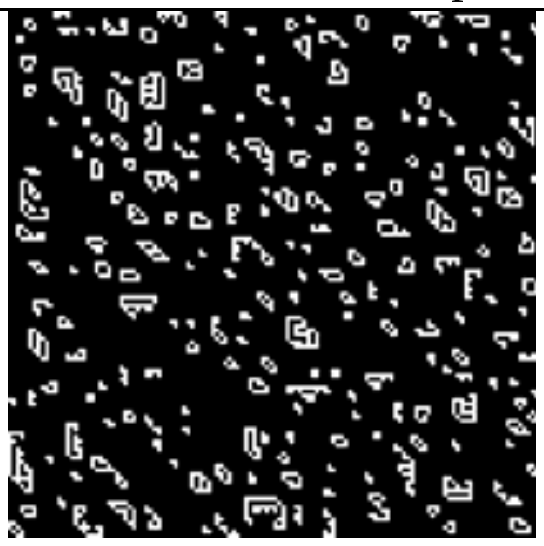<br>Seed pattern “InPattern05”.<br>$P_{death}$=0.85; $2.1\text{x}10^2$ evsteps; stable | 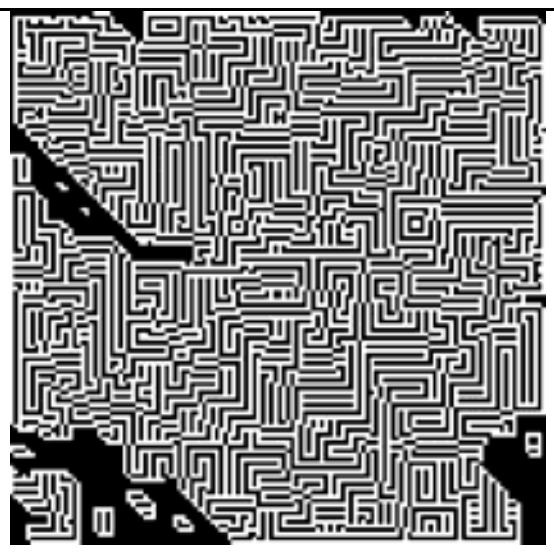<br>Seed pattern “InPattern085”.<br>$P_{death}$=0.1; $2\text{x}10^2$ evsteps;stable |

Figure 10. Two types of “ordering of chaos” dynamics for Isotropic_mask and Hex2_mask models. “Live” cells are shown white, empty cells are shown black.

However, in distinction from the standard model, maze-like patterns generated by these modified models feature only limited potentials to grow. As one can see from Figure 11 and

Figure 12, circular fragments of the stable maze-like patterns chosen as seed patterns do grow, but only until growing patterns reach a square, for isotropic_mask model, or a hexagon, for Hex2-mask model, shapes that circumscribe the shape of the seed pattern. Then the growth stops. Thus the growth is kind of “self-controlled”. In what follows we will see more examples of such a “self-controlled growth” for other models.

The limited growth capability is reflected also in the capability of the models to fill holes in seed patterns. While isotropic_mask model does fill the hole, as can be seen from Figure 11 and

Figure 12, Hex2-model fills the hole only partially leaving empty configuration with horizontal, vertical and diagonal-oriented borders similar to those of the above mentioned hexagon bounded maze-like formation.

| | | |
|---|---|---|
| 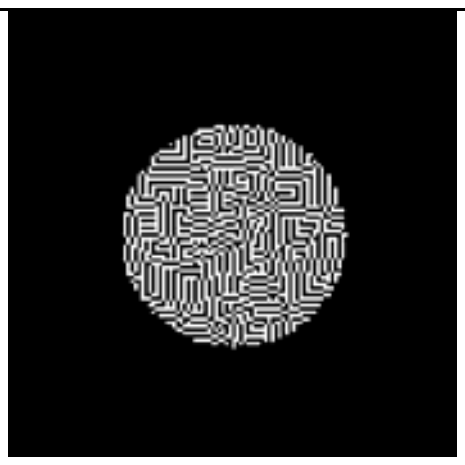<br>Circular fragment of the stable maze-like pattern ($\boldsymbol{P}_{death}$=0.4; 2x$10^2$ evsteps) used as a seed pattern. | 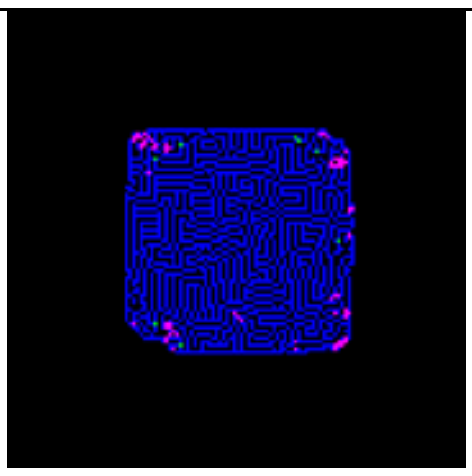<br>This fragment after 25 evsteps | 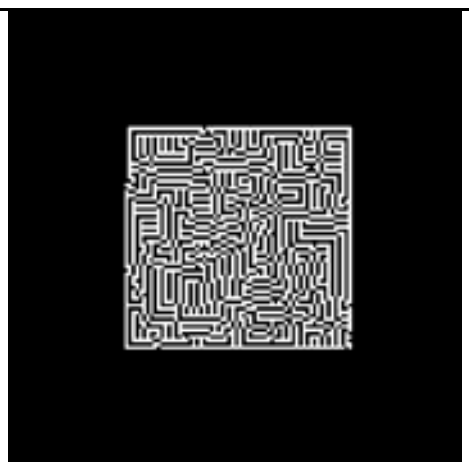<br>Same fragment stable after 2x$10^3$ evsteps. Outer square circumscribes the circular shape of the seed pattern |
| 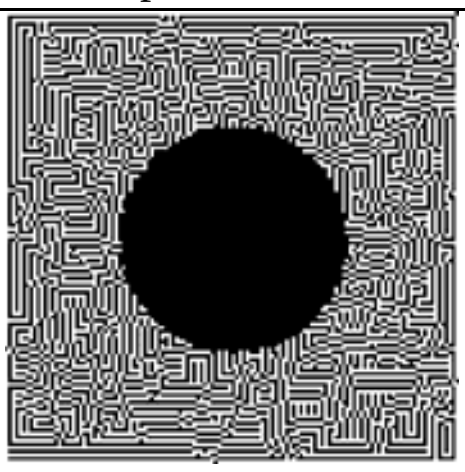<br>Circular hole un the stable maze-loke pattern ($\boldsymbol{P}_{death}$=0.4; 2x$10^2$ evsteps) used as a seed pattern. | 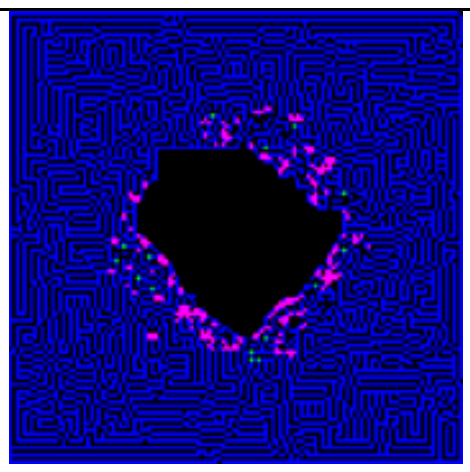<br>This hole after 25 evsteps | 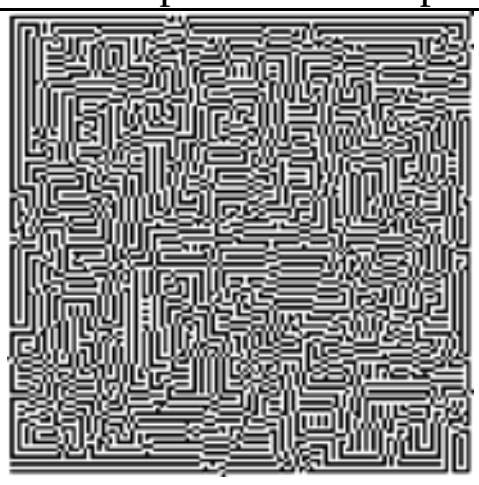<br>Stable. Hole filled after 225evol. |

Figure 11. Growth capability of maze-like patterns generated by the Isotropic_mask model. In color coded images in the middle column cells that will "die" on the next step are shown pink, cells that will give "birth" are shown green, stable cells are shown blue and empty cells are shown black. In black and white images "live" cells are shown white.

| | | |
|---|---|---|
| 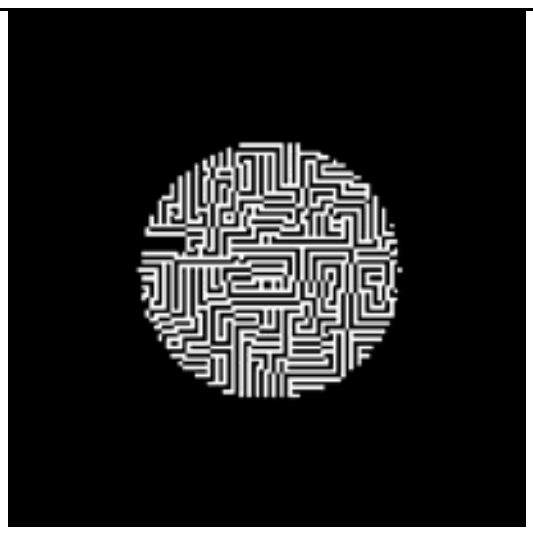<br>Circular fragment of the pattern to the left ($\boldsymbol{P}_{death}$=0.1; 2x$10^2$ evsteps) | 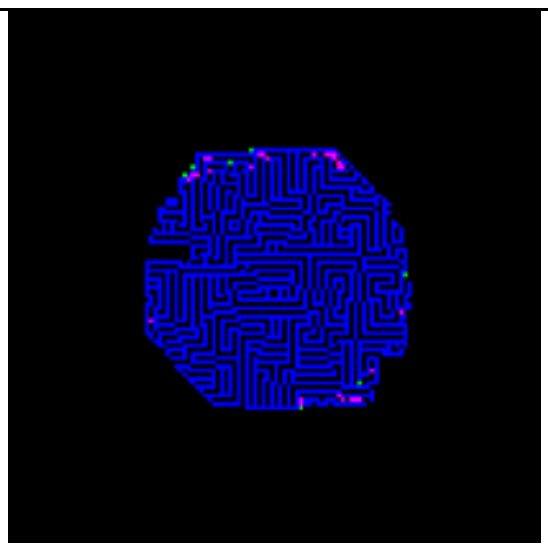<br>This fragment after 25 evsteps | 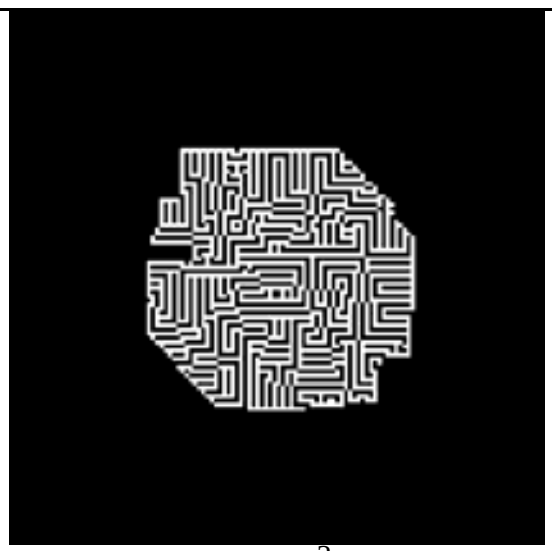<br>Stable after 1.25x$10^3$ evsteps. Outer hexagon circumscribes the circular shape of the seed pattern |
| 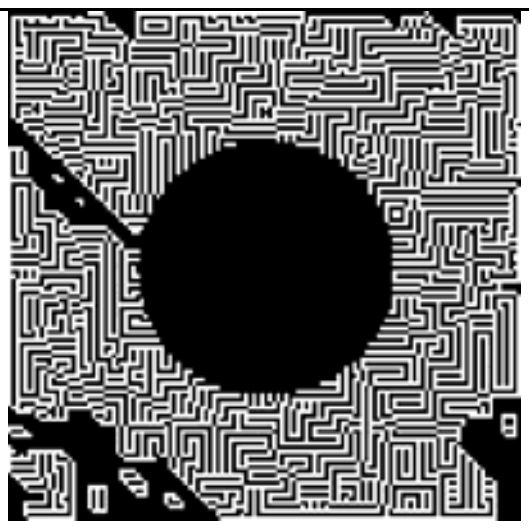<br>Circular hole in the pattern to the left ($\boldsymbol{P}_{death}$=0.1; 2x$10^2$ evsteps) | 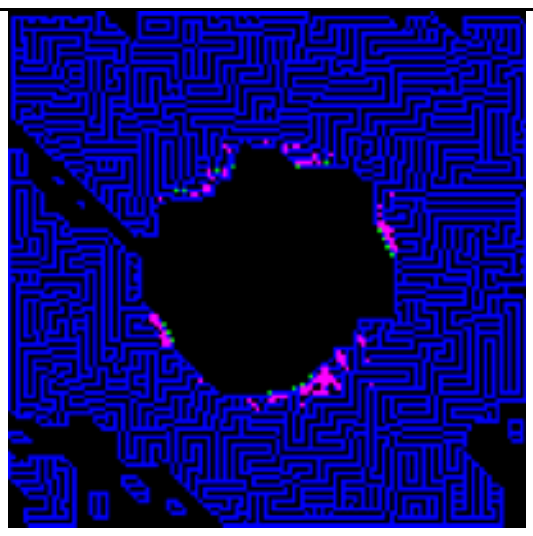<br>This hole after 25 evsteps | 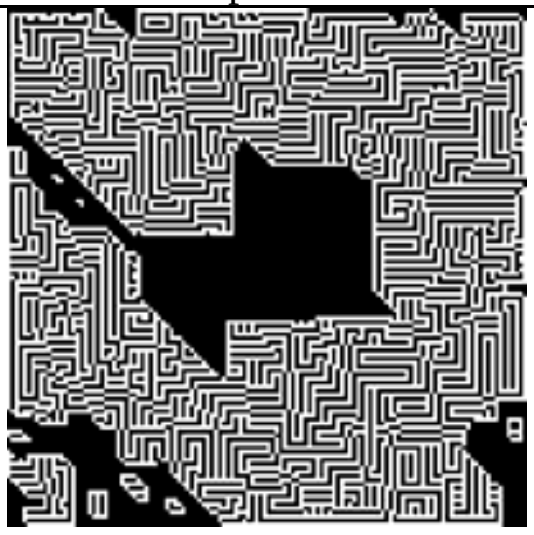<br>Stable. This hole was partially filled after 2.2x$10^2$ evsteps. |

Figure 12. Growth capability of maze-like patterns generated by the Hex2_mask model**.** In color coded images in the middle column cells that will "die" on the next step are shown pink, cells that will give "birth" are shown green, stable cells are shown blue and empty cells are shown black. In black and white images "live" cells are shown white.

Figure 13 and

Figure 14 evidence that maze-like patterns generated be Isotropic_mask as well as those of Hex2_mask model preserve certain compatibility and allow implantation of one to another, similarly to what was observed for the standard model.

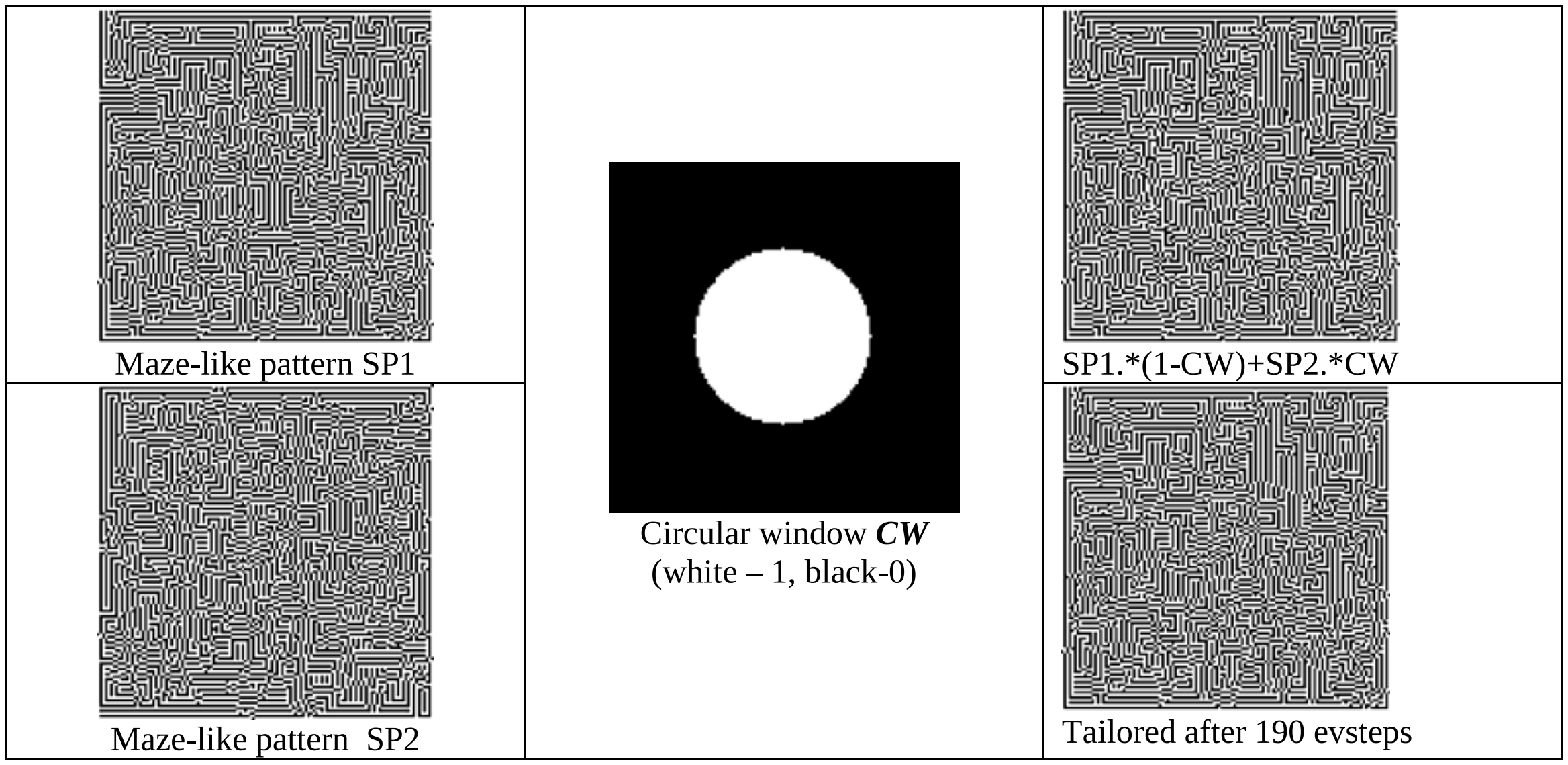

Figure 13. Isotropic_mask model: implantation of a fragment of one stripy pattern into another. Initial patterns SP1 and SP2 are shown in the left column. Pattern SP1 with implanted central circular fragment of the pattern SP2 and the resulting from it as a seed pattern the new stable maze-like pattern are shown in the right column. Circular mask used for extracting the implanted pattern is shown in the middle.

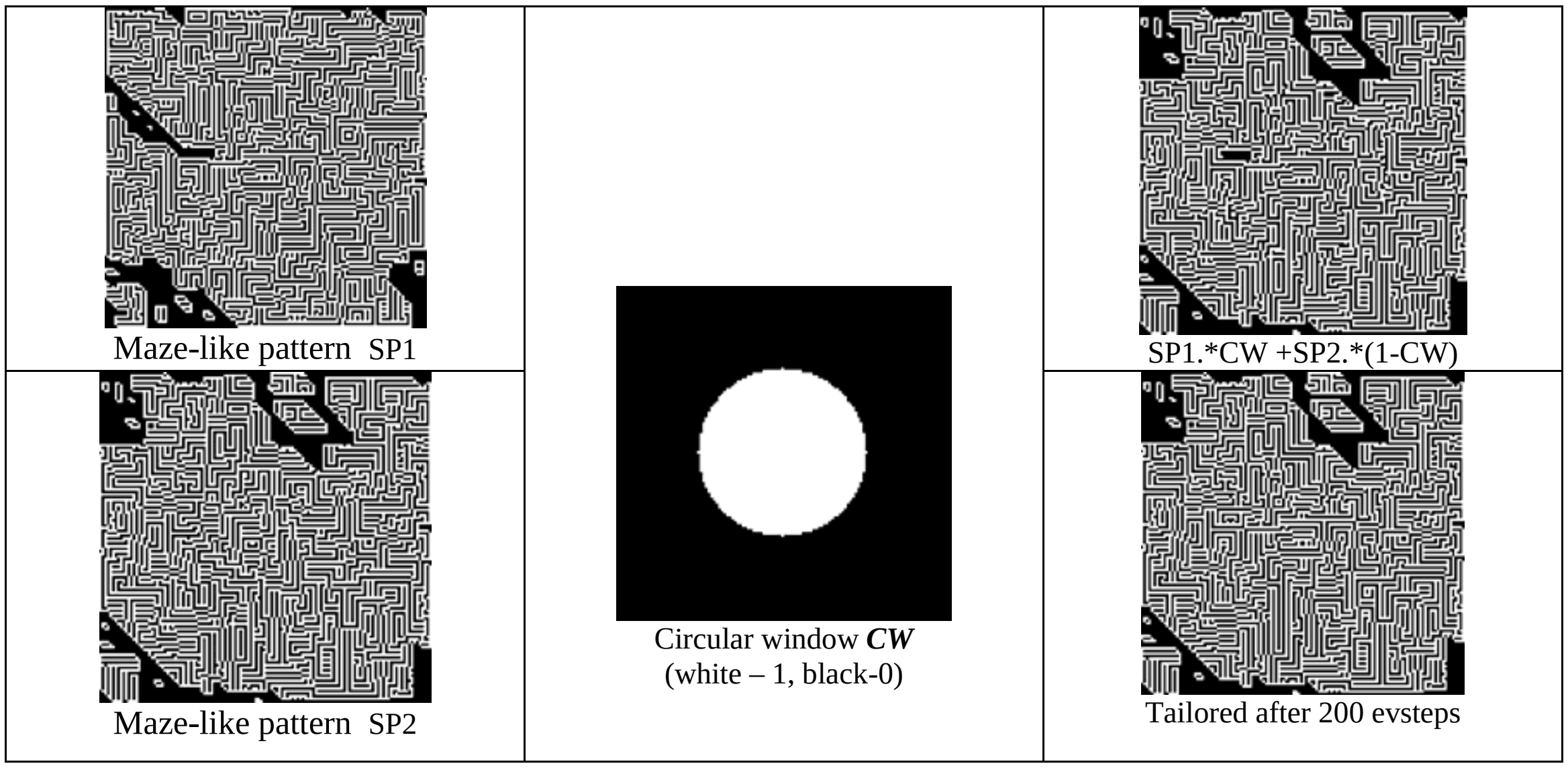

Figure 14. Hex2_mask model: implantation of a fragment of one maze-like pattern into another. Initial patterns SP1 and SP2 are shown in the left column. Pattern SP1 with implanted central circular fragment of the pattern SP2 used as a seed pattern and the resulting from new stable maze-like pattern are shown in the right column. Circular mask used for extracting the implanted pattern is shown in the middle.

Figure 15 provides further illustrations of the “self-controlled growth” property of the Isotropic-mask and Hex2_mask models.

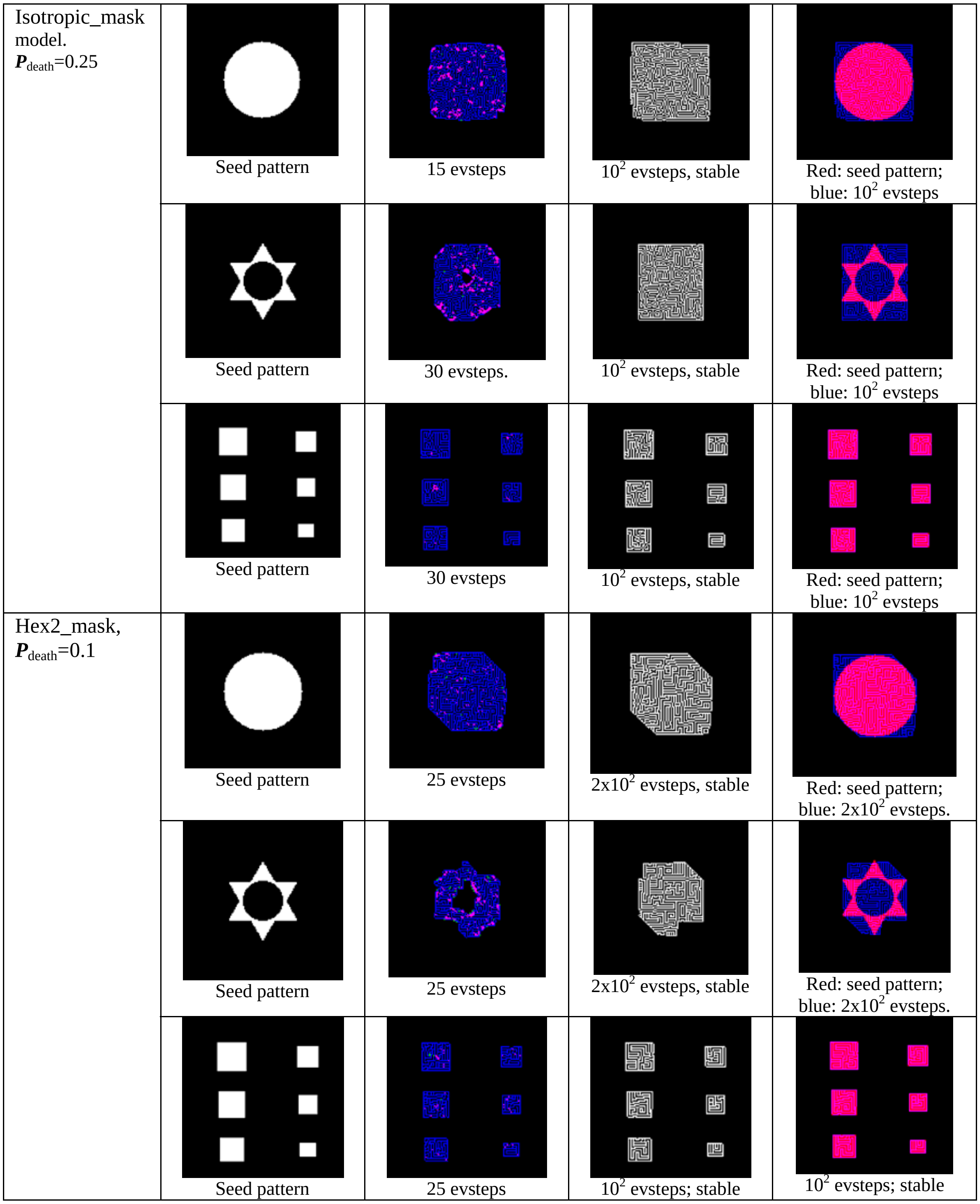

Figure 15. Examples of the evolution of the Hex2_mask model from solid seed patterns. In color coded images in the second column cells that will “die” on the next step are shown pink, cells that will give “birth” are shown green, stable cells are shown blue and empty cells are shown black. In black and white images “live” cells are shown white.

Models with “Cross4 and “Cross4diag” masks do not produce maze-like patterns. Rather their fixed points are, for the entire range of the probability of “death” $\mathbf{0 < \mathit{P}_{death} \leq 1}$, what can be called “Manhattan-like” patterns (
Figure 16).

| Mask “Cross4” Seed pattern “Inpattern05” | $\mathbf{0 < \mathit{P}_{death} \leq 1}$ | | |
|---|---|---|---|
| | 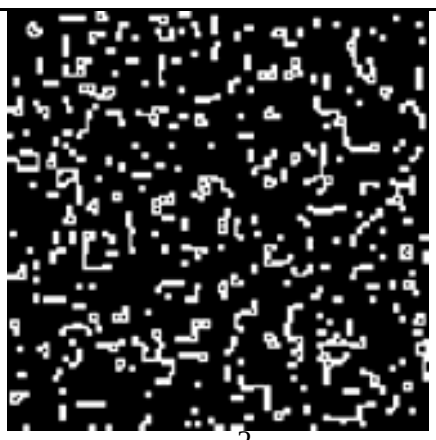<br>$\mathbf{\mathit{P}}_{\text{death}}$=1; $10^2$ evsteps.<br>Stable with 3 oscillators | 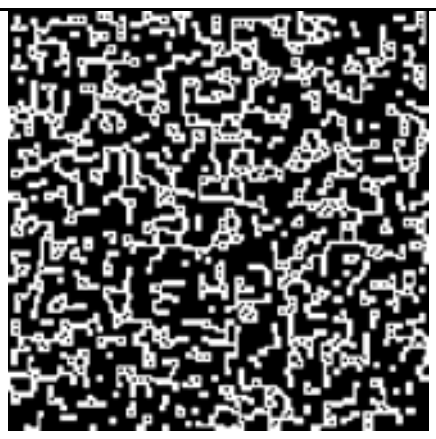<br>$\mathbf{\mathit{P}}_{\text{death}}$=0.5; 80 evsteps. Stable | 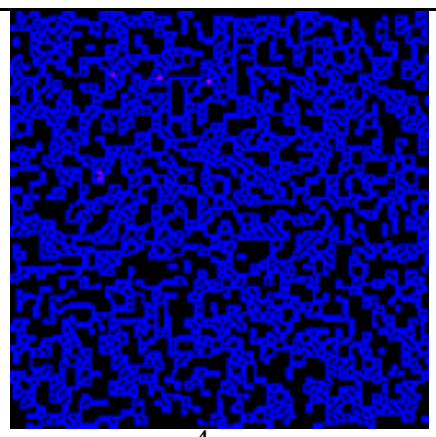<br>$\mathbf{\mathit{P}}_{\text{death}}$=0.1; $10^4$ evsteps. Stable (blue) with oscillators (pink pixels) |
| Mask “Cross4” Seed patterns “ Circle” and “Star” | 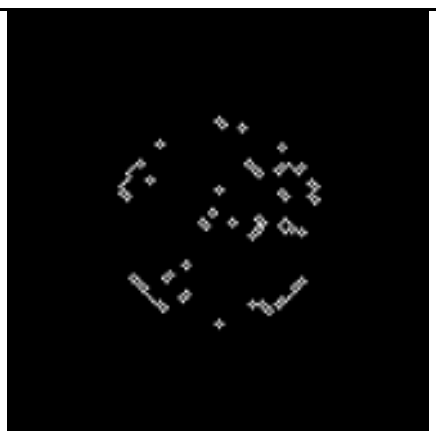<br>$\mathbf{\mathit{P}}_{\text{death}}$=0.75; 25 evsteps. Stable (for $\mathbf{\mathit{P}}_{\text{death}}$=1, all cells die out ) | 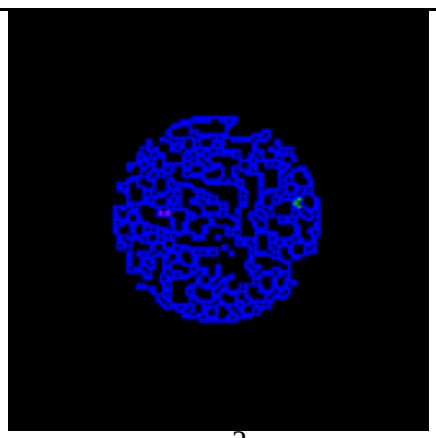<br>$\mathbf{\mathit{P}}_{\text{death}}$=0.25; $4\text{x}10^2$ evsteps Stable with two oscillators (green and pink pixels) | 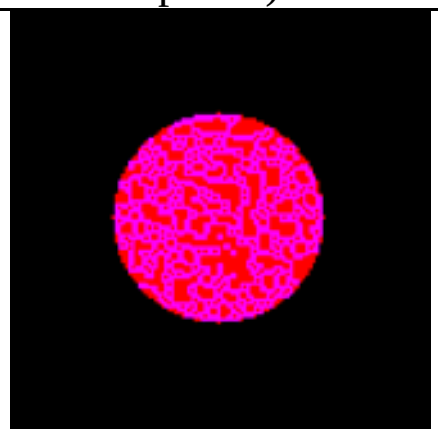<br>Red: seed pattern;<br>blue: $\mathbf{\mathit{P}}_{\text{death}}$=0.25; $4\text{x}10^2$ evsteps |
| | 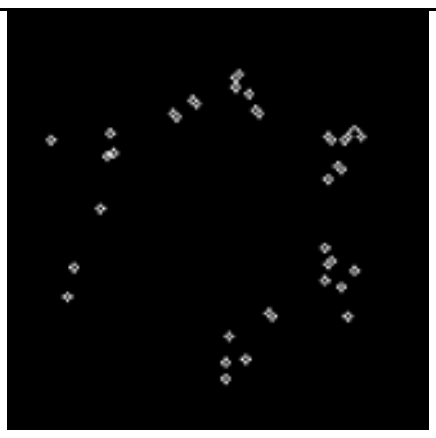<br>$\mathbf{\mathit{P}}_{\text{death}}$=0.75; 25 evsteps. Stable | 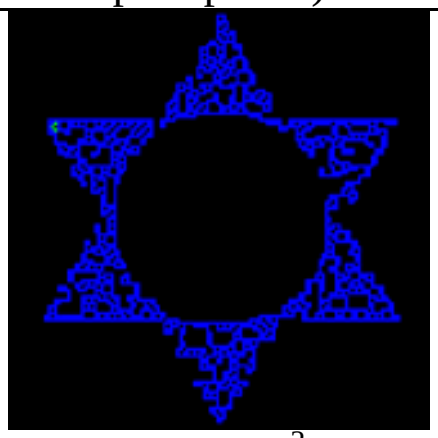<br>$\mathbf{\mathit{P}}_{\text{death}}$=0.25; $2\text{x}10^2$ evsteps.<br>Stable with two oscillators (green and pink pixels) | 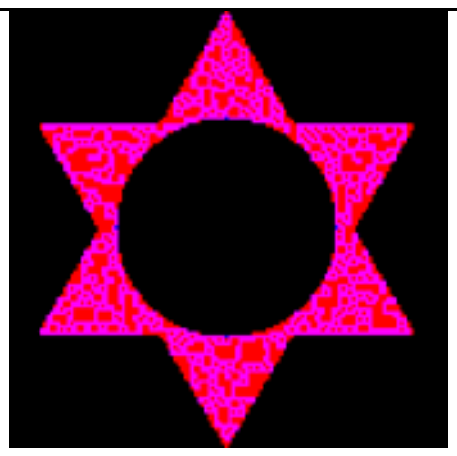<br>Red: seed pattern;<br>blue: $2\text{x}10^2$ evsteps |
| Mask “Cross4diag”; Seed pattern “Inpattern05” | 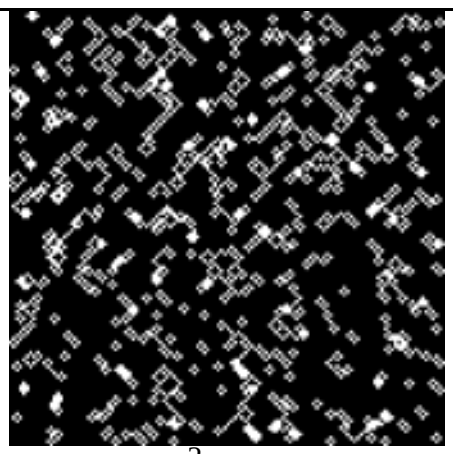<br>$\mathbf{\mathit{P}}_{\text{death}}$=1; $10^2$ evol. evstepss.<br>Stable with 1 oscillator | 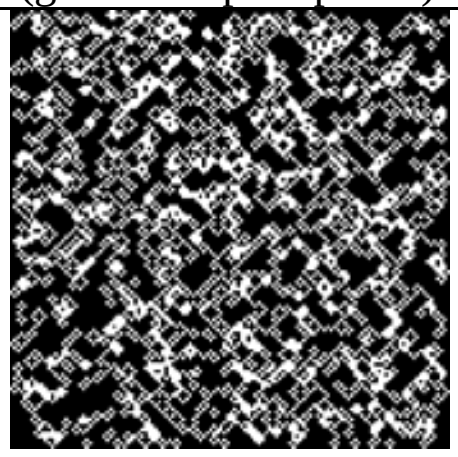<br>$\mathbf{\mathit{P}}_{\text{death}}$=0.5; 50 evol. evstepss.<br>Stable with two oscillators | 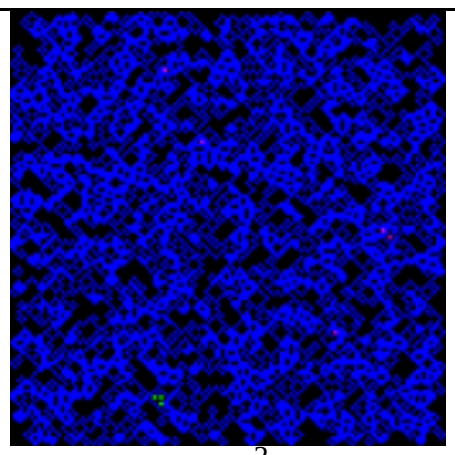<br>$\mathbf{\mathit{P}}_{\text{death}}$=0.1; $1.5\text{x}10^3$ evol. evsteps.<br>Stable (blue) with 5 oscillators (red and green) |

Figure 16. Manhattan-type stable patterns generated by the model with masks “ Cross4” and “Cross4diag”. In color coded images (except two last images in the second and third rows), cells that will “die” on the next step are shown pink, cells that will give “birth” are shown green, stable cells are shown blue and empty cells are shown black. In black and white images “live” cells are shown white.

## 5. *“Coherent shrinkage” and “eternal life in a bounded space” dynamics modes*

Perhaps, the most amazing types of evolutionary dynamics observed in the experiments are, along with above-mentioned “self-controlled growth”, “coherent shrinkage” and “eternal life in a bounded space”. They were observed with some models for the probabilities of “death” in the middle of the range 0÷1.

Figure 17 and Figure 18 illustrate the phenomenon of “Coherent shrinkage” observed for a dense chaotic seed pattern “InPattern05”. The phenomena of “self-controlled growth” and “coherent shrinkage” can be comprehended even better using Figure 19, Figure 20, Figure 21, Figure 22, Figure 23 and Figure 24 which illustrate evolution of “solid” seed patterns.

| **”Isotropic_mask”, seed pattern “InPattern05”** $P_{death}$ **= 0.8** | | | |
|---|---|---|---|
| 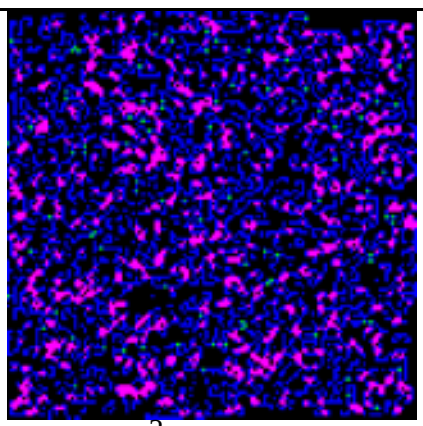<br>$10^3$ evsteps. | 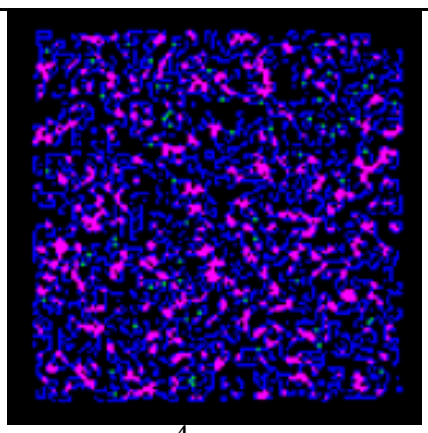<br>$5x10^4$ evsteps. | 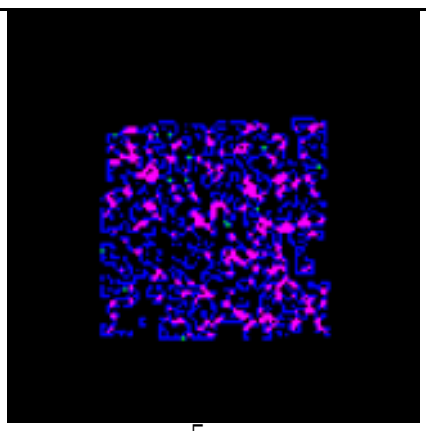<br>$1.5x10^5$ evsteps. | 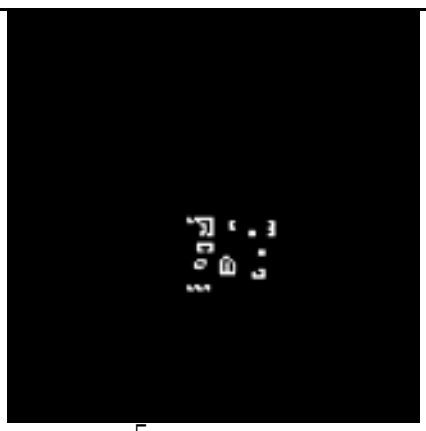<br><$2x\ 10^5$ evsteps, stable |
| **Diagonal_mask model, seed pattern “InPattern05”,** $P_{death}$=0.75 | | | |
| 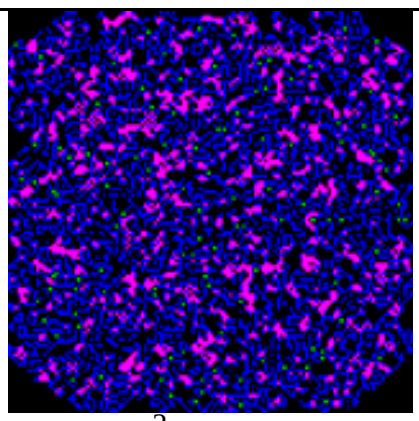<br>$10^2$ evsteps | 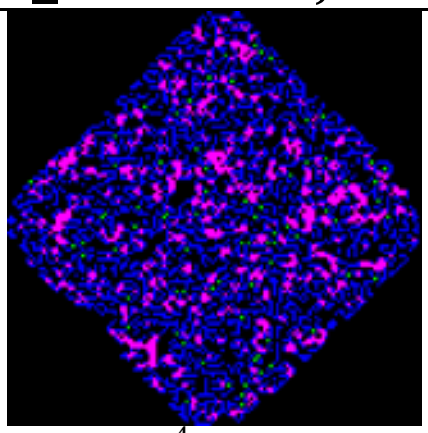<br>$5x10^4$ evsteps; | 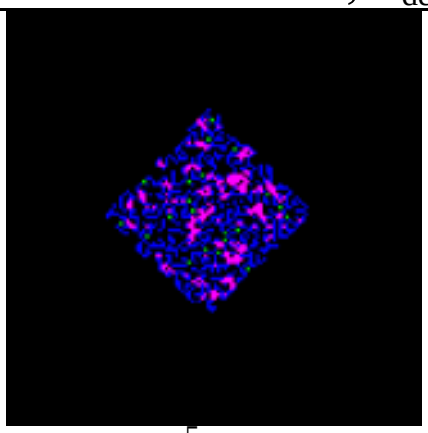<br>$1.3x10^5$ evsteps. | 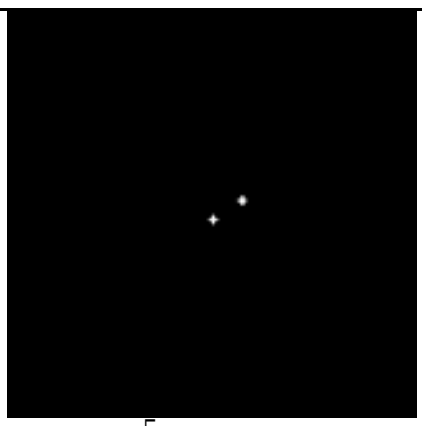<br><$1.5x10^5$ evsteps, stable |
| **Cross_model, seed pattern “InPattern05”;** $P_{death}$=0.55; | | | |
| 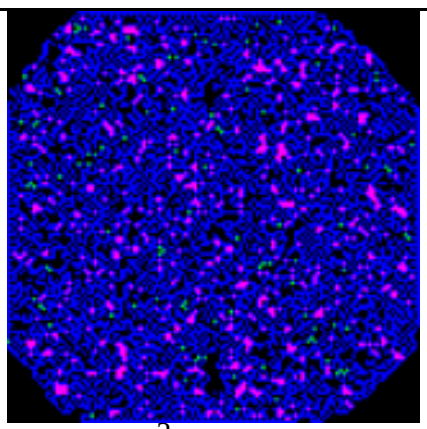<br>$10^3$ evsteps | 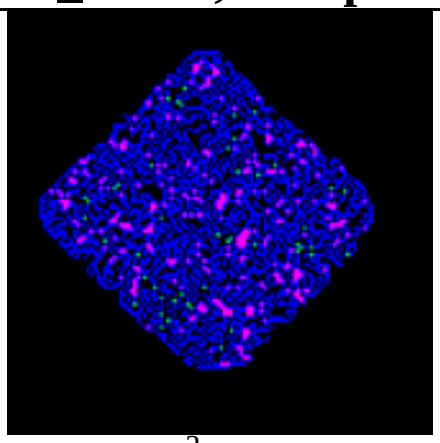<br>$5x10^3$ evsteps | 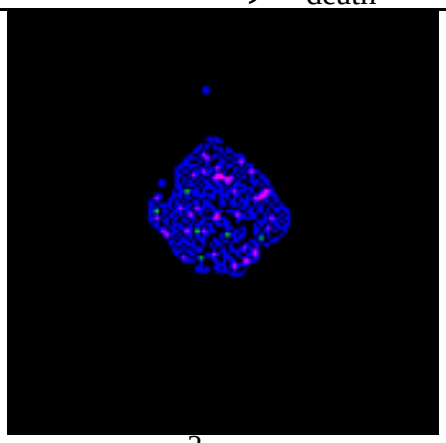<br>$7x10^3$ evsteps | 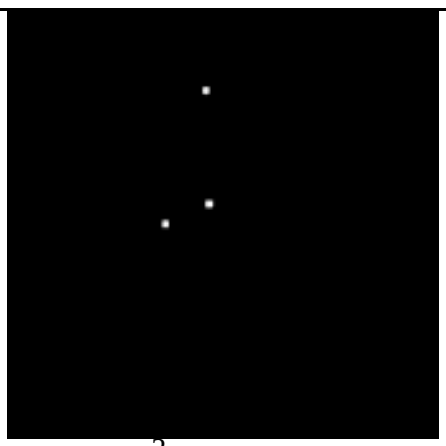<br>$7.75x10^3$ evsteps, stable |

Figure 17. “Coherent shrinkage” of patterns emerged from seed pattern “InPattern05” for Isotropic_mask, Diagonal_mask and Cross_mask models. In color coded images, cells that will “die” on the next step are shown pink, cells that will give “birth” are shown green, stable cells are shown blue and empty cells are shown black. In last images of every row, “live” cells are shown white.

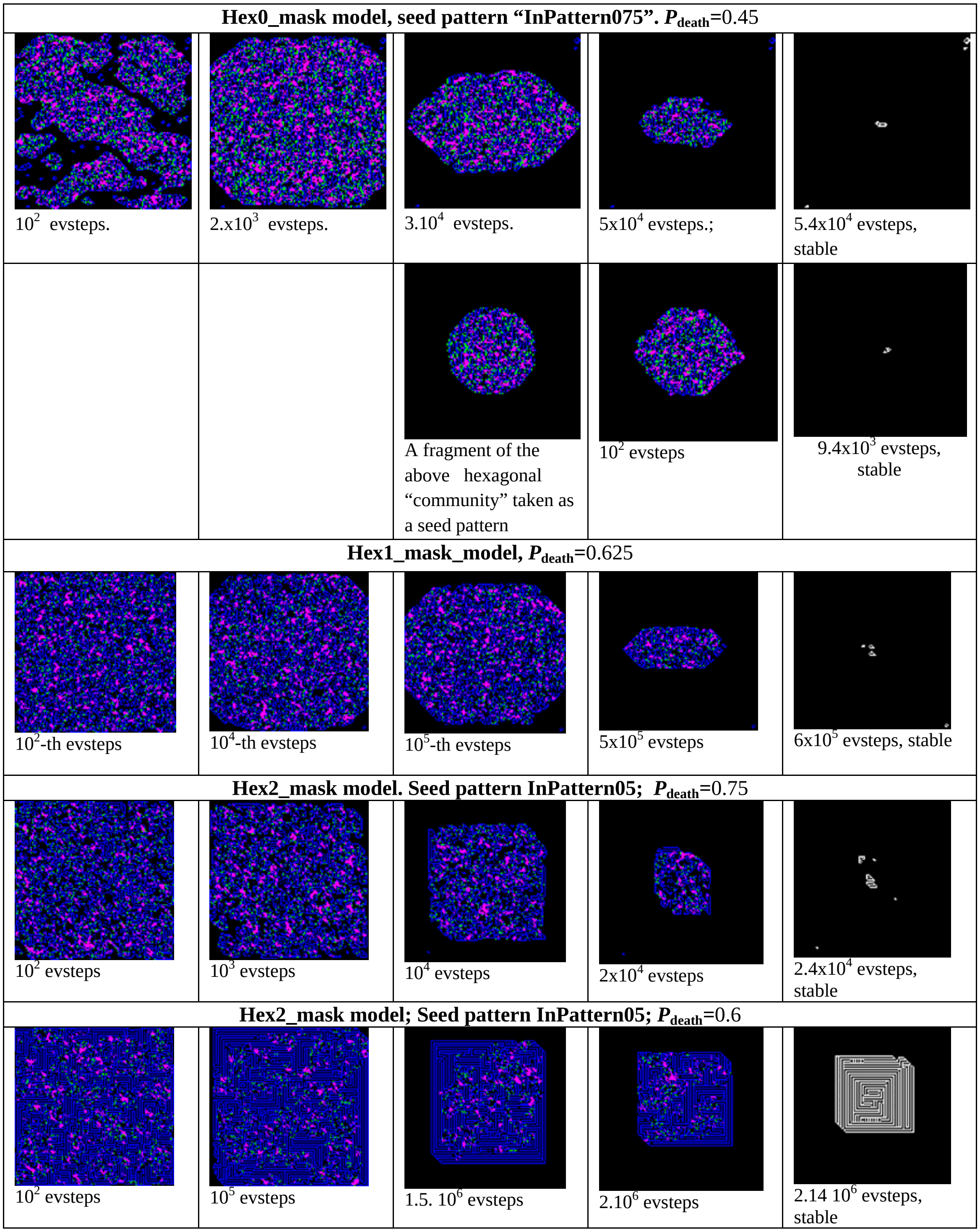

**Figure 18.** “Coherent shrinkage” of patterns emerged from seed pattern “InPattern05” for Hex0-, Hex1- and Hex2_mask models. Images in the second from the top row show evolution of a fragment of the pattern obtained after $3.10^4$ evsteps planted in an empty space. In color coded images cells that will “die” on the next step are shown pink, cells that will give “birth” are shown green, stable cells are shown blue and empty cells are shown black. In last images of every row, “live” cells are shown white.

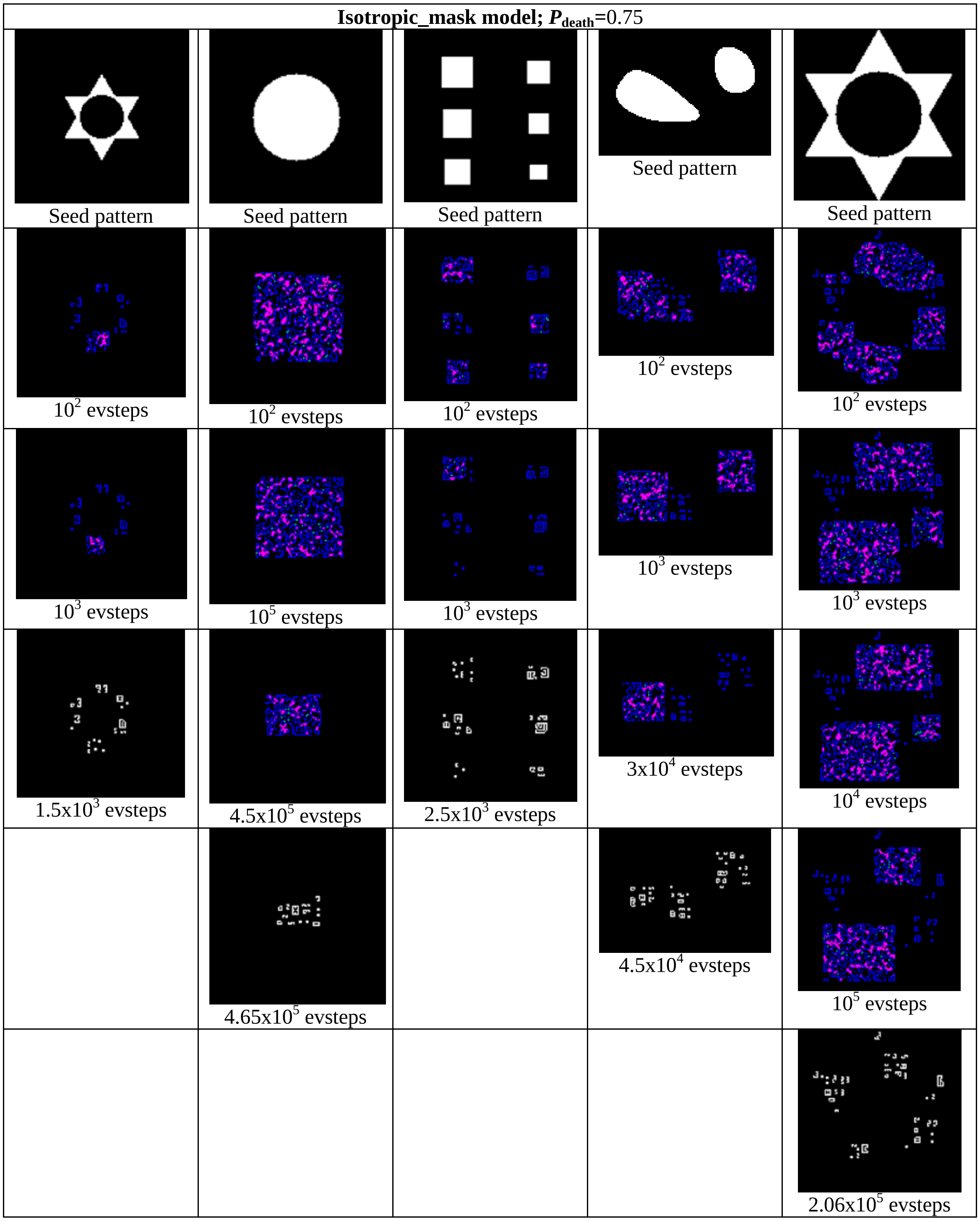

Figure 19. “Self-controlled growth” and “Coherent shrinkage” of patterns emerged from solid seed patterns for Isotropic_mask. In color coded images cells that will “die” on the next step are shown pink, cells that will give “birth” are shown green, stable cells are shown blue and empty cells are shown black. In black and white images “live” cells are shown white.

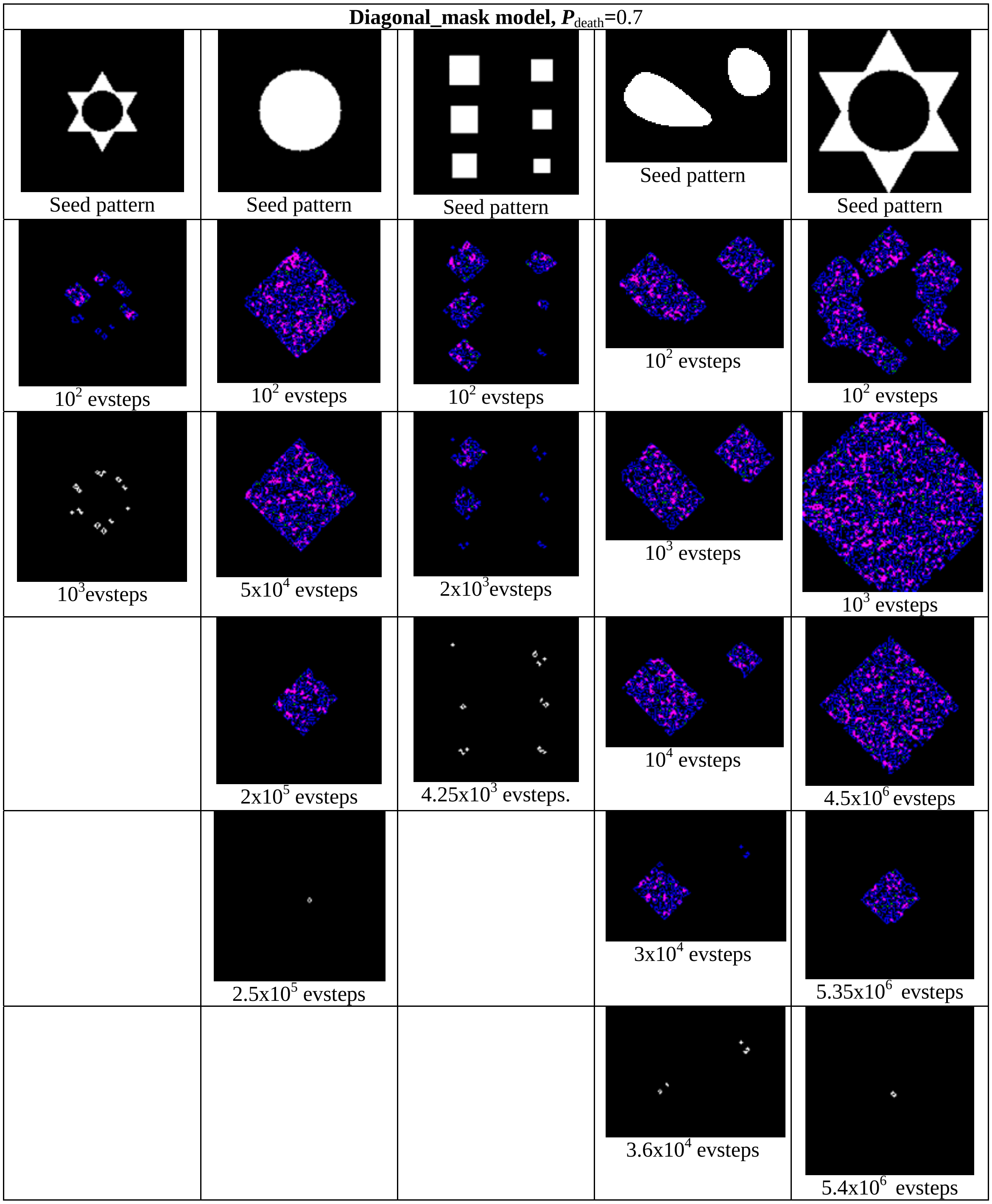

**Figure 20.** "Self-controlled growth" and Coherent shrinkage" of patterns emerged from solid seed patterns for the Diagonal_mask model. In color coded images cells that will "die" on the next step are shown pink, cells that will give "birth" are shown green, stable cells are shown blue and empty cells are shown black. In last images of every column, "live" cells are shown white.

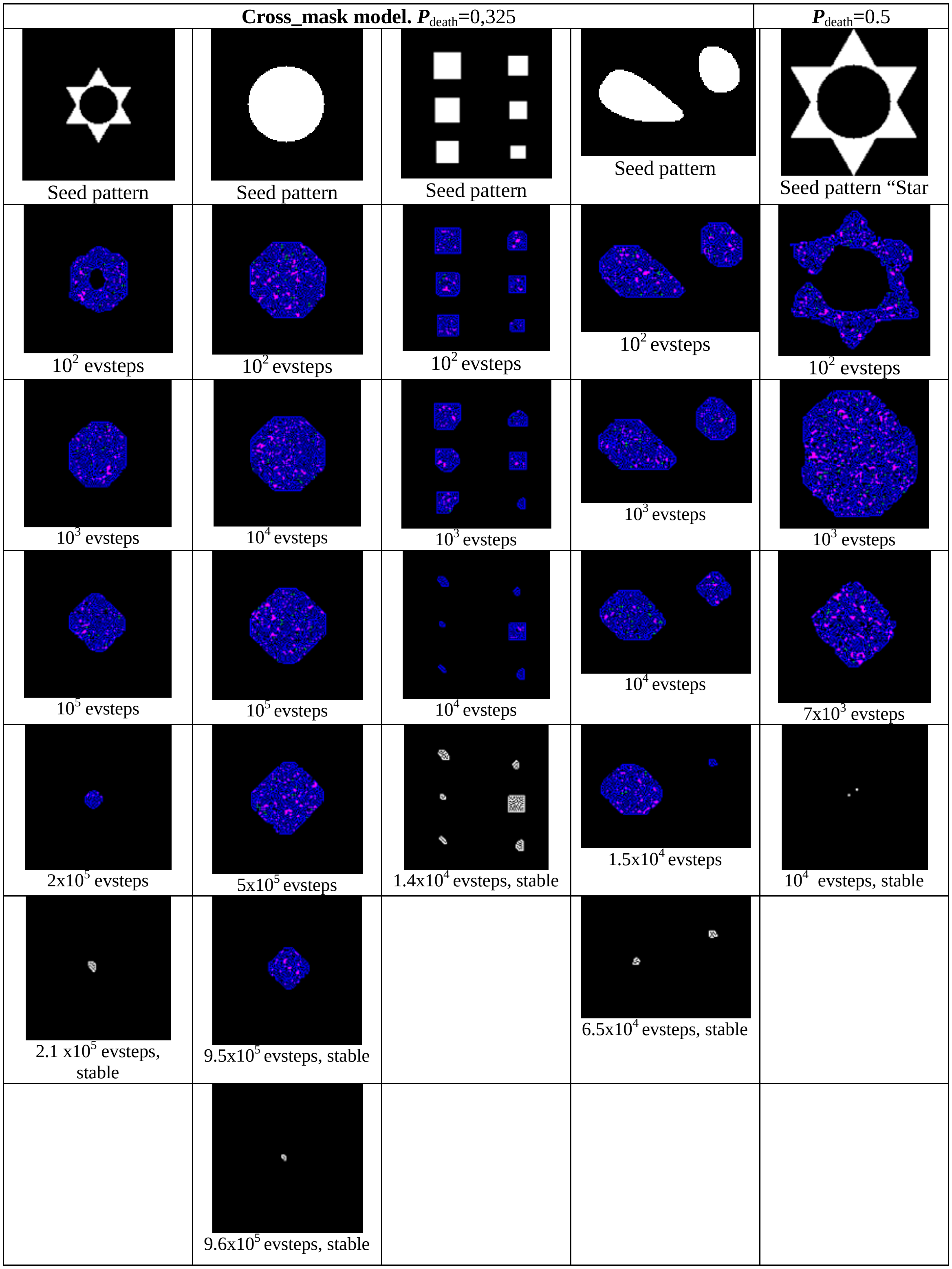

Figure 21. "Self-controlled growth" and Coherent shrinkage" of patterns emerged from solid seed patterns for the Cross_mask model**.** In color coded images cells that will "die" on the next step are shown pink, cells that will give "birth" are shown green, stable cells are shown blue and empty cells are shown black**.** In last images of every column, "live" cells are shown white.

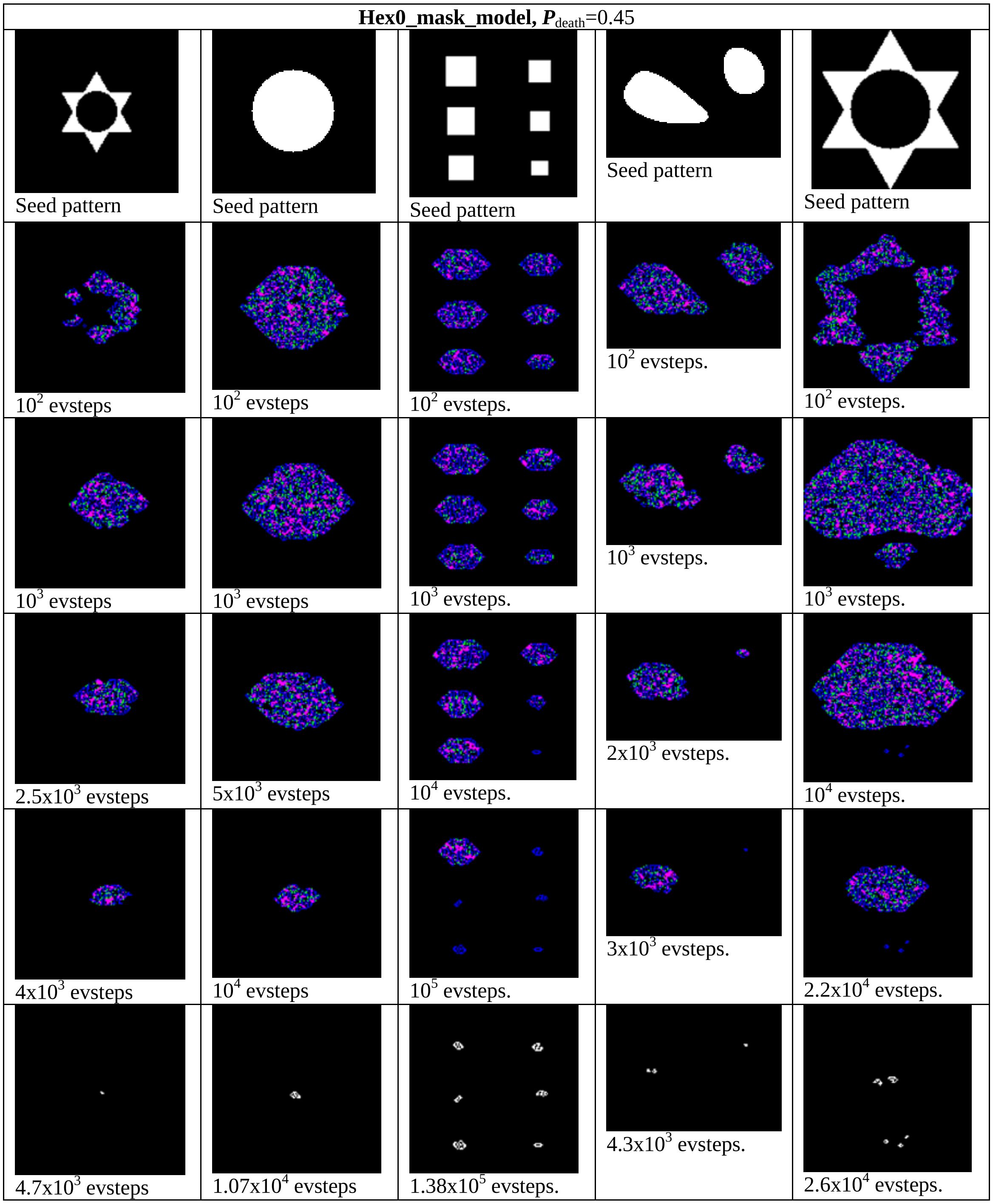

Figure 22. “Self-controlled growth” and Coherent shrinkage” of patterns emerged from solid seed patterns for the Hex0_mask model. In color coded images cells that will “die” on the next step are shown pink, cells that will give “birth” are shown green, stable cells are shown blue and “empty” cells arte shown black. In black and white images “live” cells are shown white.

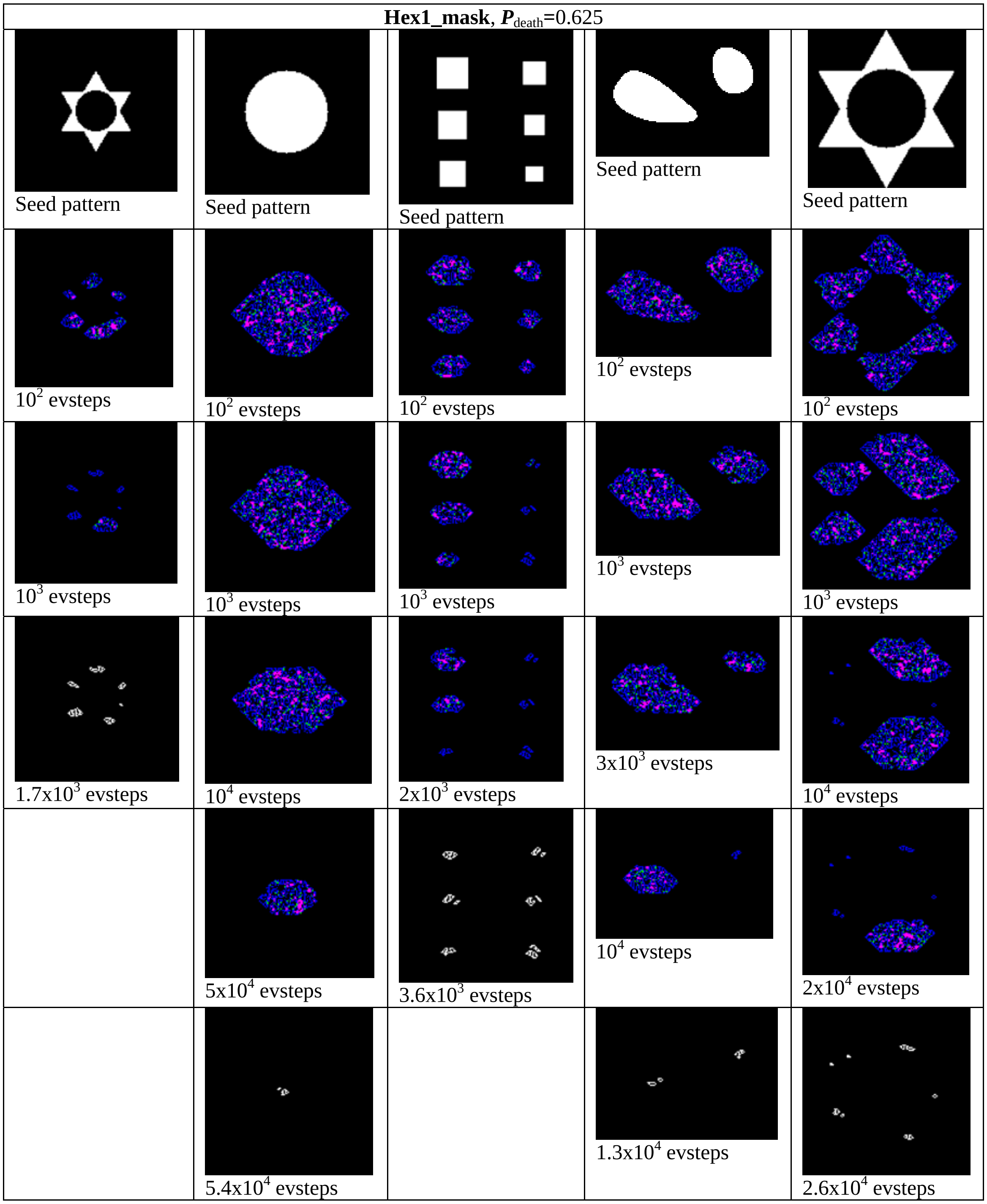

Figure 23. “Self-controlled growth” and “Coherent shrinkage” of patterns emerged from solid seed patterns for the Hex1_mask model. In color coded images cells that will “die” on the next step are shown pink, cells that will give “birth” are shown green, stable cells are shown blue and “empty” cells arte shown black. In black and white images “live” cells are shown white.

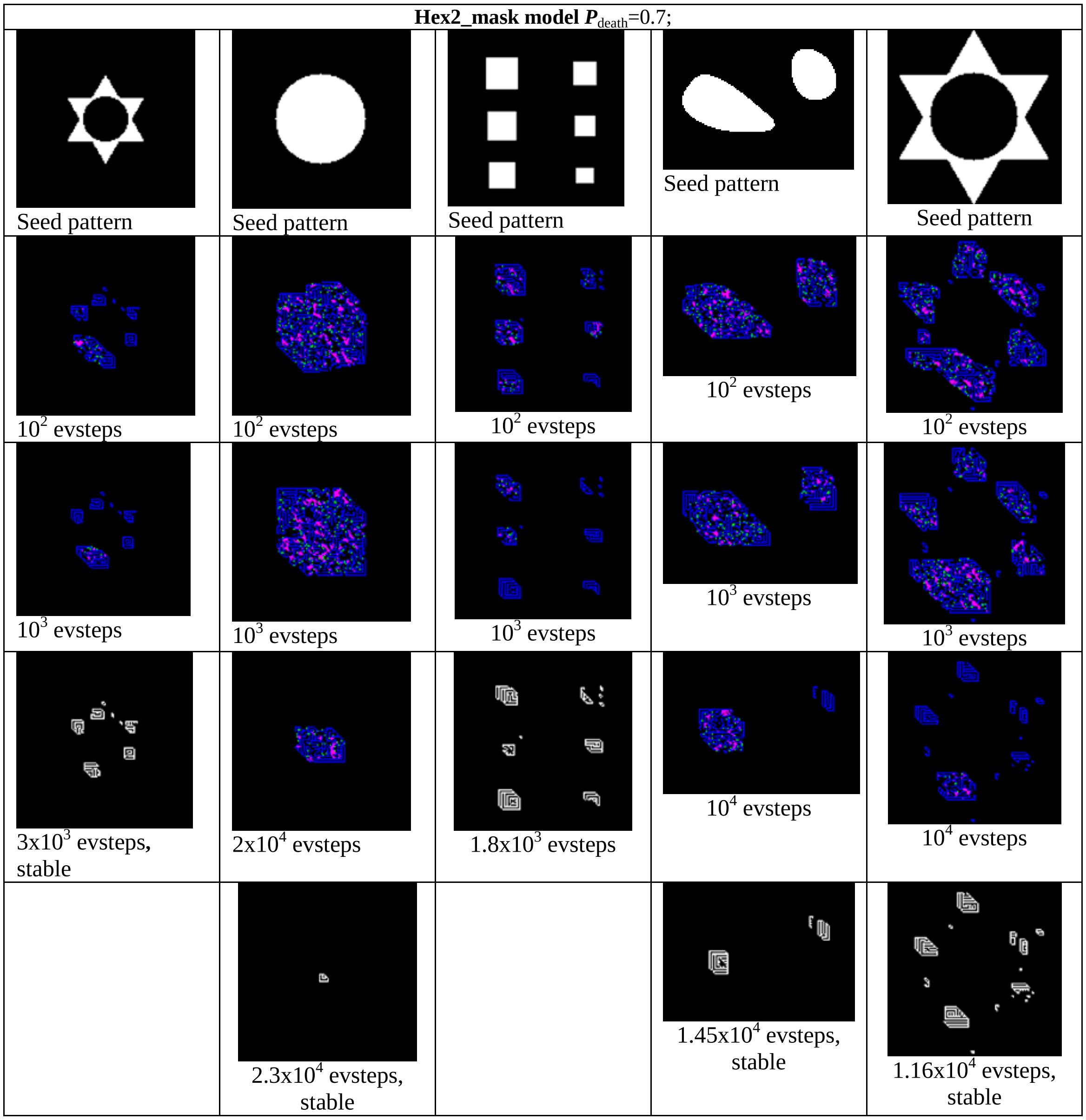

Figure 24. “Self-controlled growth” and Coherent shrinkage” of patterns emerged from solid seed patterns for the Hex2_mask model. In color coded images cells that will “die” on the next step are shown pink, cells that will give “birth” are shown green, stable cells are shown blue, “empty” cells arte shown black. In black and white images “live” cells are shown white.

As one can see from the figures, in the “coherent shrinkage” mode the models pass, in course of evolution, through stages of a sort of entire “life cycle”:

- “Birth”: loci of growth emerge in seed patterns.
- “Childhood and adolescence”: born formations grow in size, forming kind of “communities” of cells. For chaotic seed patterns, in which “live” cells fill more all less uniformly the entire “vital space”, this growth goes on within the “vital space”. For “solid” seed patterns, the growth is “self-controlled”: it goes on till “communities” reach a shape bounded, depending on the model, by a square (Isotropic and Diagonal_mask models), an octagon (Cross_mask model) or a hexagon (Hex0, Hex1 and Hex2-mask models). The emerged shaped communities stop growing further unless they touch another neighbor “community”. In this case touching communities merge to form larger “communities”, which continue growing till they reach a similarly bounded shape of a larger size. In such a way “communities” reach a state of “maturity”.

- The state of “maturity”: bounded shaped mature “communities” stay like islands in the “ocean” of empty cells and keep their activity (“births” and “deaths”) and their overall size and shape during a certain number of evolution steps, which depends on the probability of death: the lower the probability of death the larger this “population” stability period.
- “Senescence”. After a certain period of relative stability in size of their populations, “communities” begin to gradually shrink. The shrinkage appears to be “coherent”: the “communities” are coherently shrinking from their borders preserving isomorphism of their shapes till the very end, when they either completely disintegrate to nil or, most frequently, end up with one of stable formations. The speed of the shrinkage depends on the probability of death and of the “community” size: the lower the probability of death and the larger the “community” size the lower the shrinkage speed. Some experimental data on the number of evolution steps from the beginning of growth to reaching a stable point obtained for the Hex0_mask model and the probability of “death” $\boldsymbol{P_{death} = 0.25}$ with seed patterns “Square” of different size are presented in

  Figure 25 along with their analytical approximation as a fourth-power function of the “shaped community” size in certain normalized units.

It is remarkable that, as one can see in the lower row of Figure 17, “communities”, in the course of the “coherent shrinkage”, preserve a capability of growth: if one extracts a fragment of a shrinking “community” and plants it into an empty space, the planted fragment resumes growing until it reaches a maturity state in a bounded shape, characteristic for the given model; after that it starts shrinking in the same way as its "mother community" does.

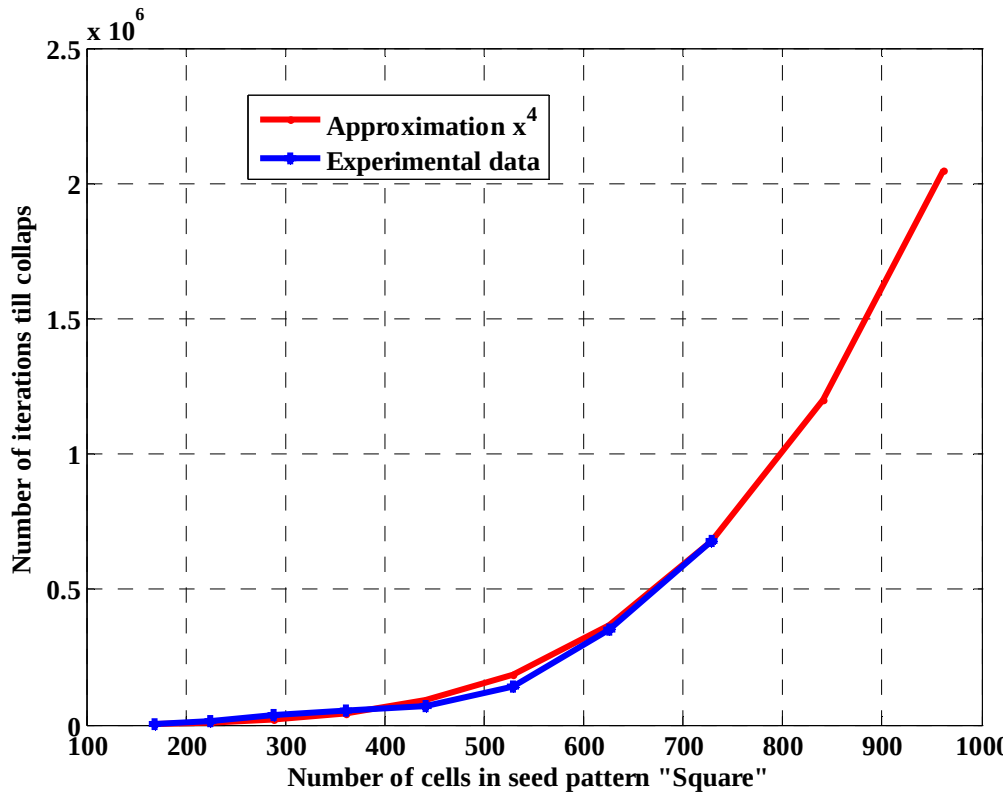

Figure 25. Experimental data on the number of evolution steps from birth to collapse for the Hex0_mask model with $\boldsymbol{P_{death}} = \mathbf{0.25}$ and seed patterns in form of squares of different sizes and their numerical approximation

As we have already indicated, "coherent shrinkage" slows down with decreasing the probability of "death". For the standard, Isotropic_mask, "Cross_mask", Hex0_mask, and Hex1_mask models, this slowing down might be so substantial that the dynamics of the models appears as "eternal life": upon reaching the state of "maturity", communities stay active (in terms of "births" and "deaths") and keep their outer bounds during millions of evolution steps. We illustrate this in Figure 27, Figure 28 and Figure 28 for "chaotic" seed patterns and in Figure **30,** Figure **31,** Figure **32**, Figure 33, Figure **34** and Figure 35 for solid seed patterns.

| **Standard mask model. Seed pattern "InPattern05"** $\approx \mathbf{0.3} < P_{death} < 1-\varepsilon$ $\varepsilon$ **- is an arbitrarily small number.** | | | |
|---|---|---|---|
| 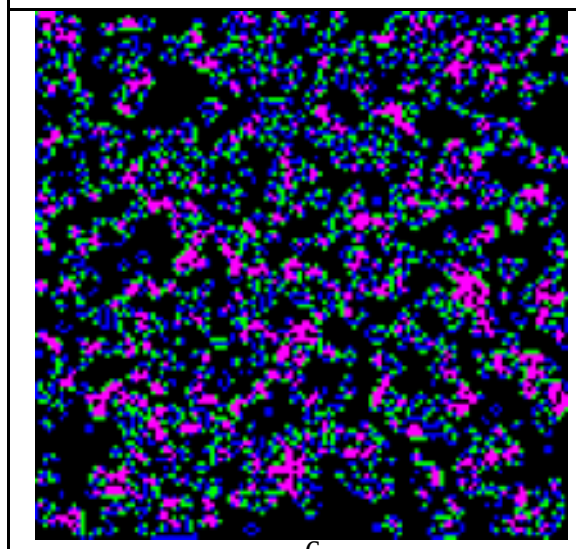 $P_{death}$=0.9; $10^6$ evsteps | 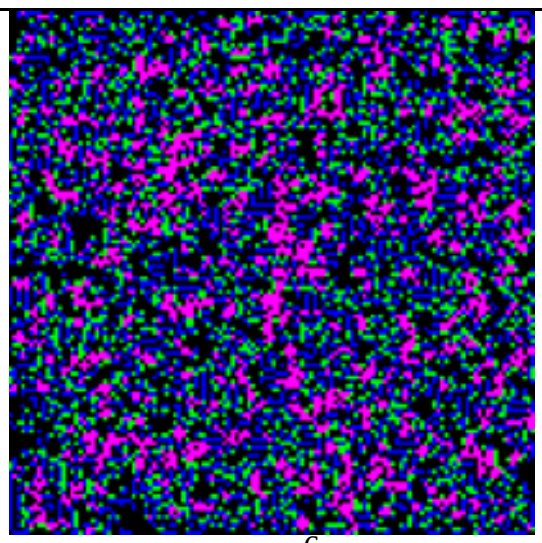 $P_{death}$=0.73; $10^6$ evsteps | 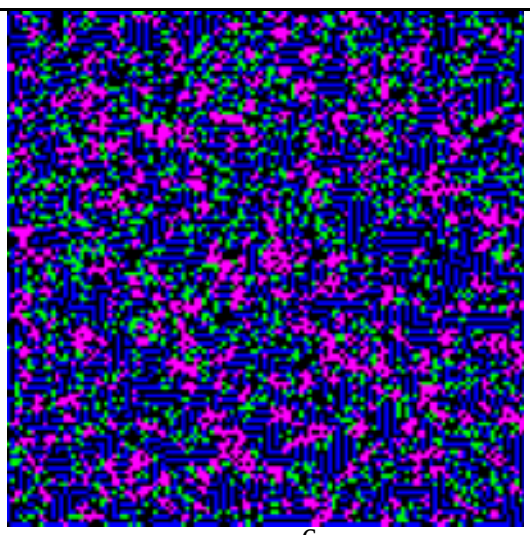 $P_{death}$=0.57; $10^6$ evsteps | 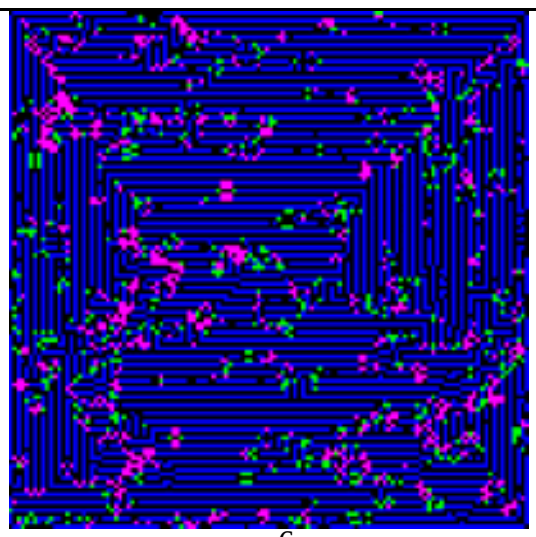 $P_{death}$=0.4; $10^6$ evsteps |
| **Isotropic_mask model. Seed pattern "InPattern09";** $\approx \mathbf{0.75} < P_{death} < \mathbf{0.5}$ | | | |

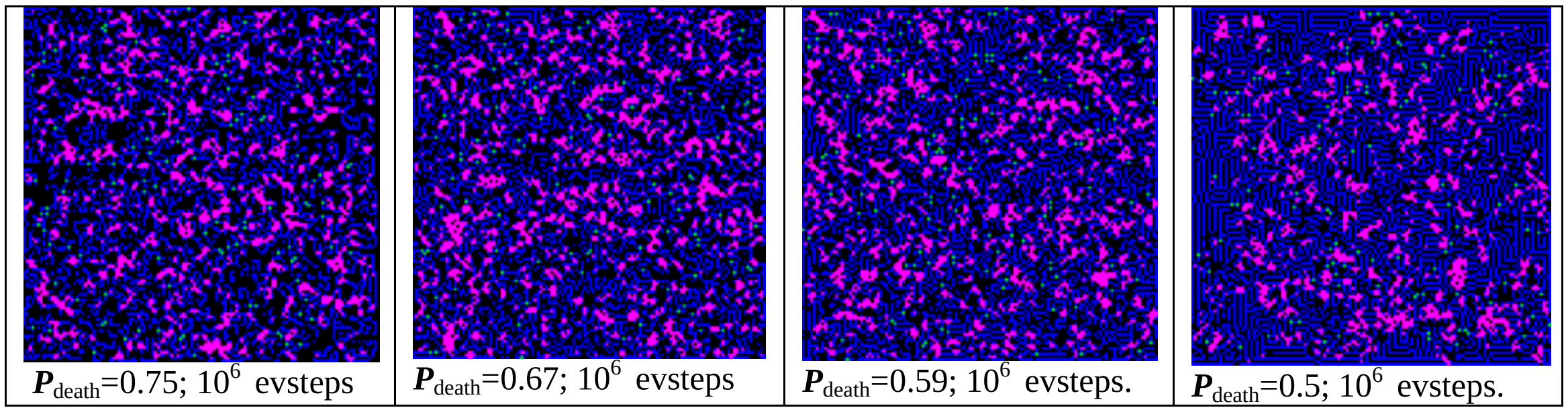

Figure 26. Standard_mask and Isotropic_mask models: "Eternal life" dynamics in the limits of the "vital space"**.** Cells that will "die" on the next step are shown pink, cells that will give "birth" are shown green and stable cells are shown blue; "empty" cells are shown black.

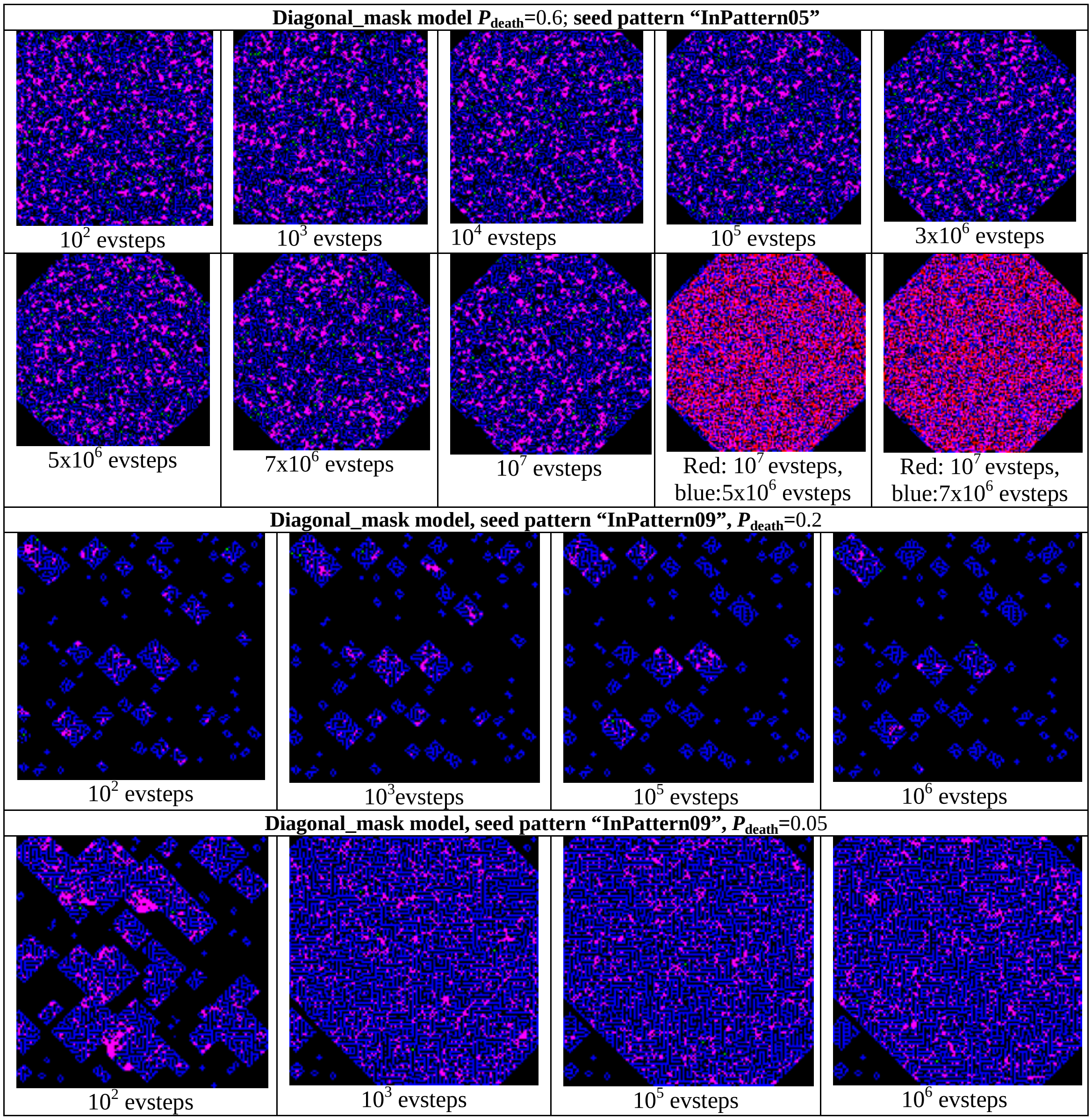

Figure 27. Diagonal_mask models: “eternal life” in a bounded space for various probabilities of “death” and “chaotic” seed patterns with densities of “live” cells 50% (first two rows) and 10% (last two rows). Last two images in the second row are color-coded, as it is indicated in the caption, in order to demonstrate that the outer shape of the octagon does not shrink noticeably after $5 \times 10^6$ evolution steps. In other color coded images cells that will “die” on the next step are shown pink, cells that will give “birth” are shown green and stable cells are shown blue; “empty” cells are shown black.

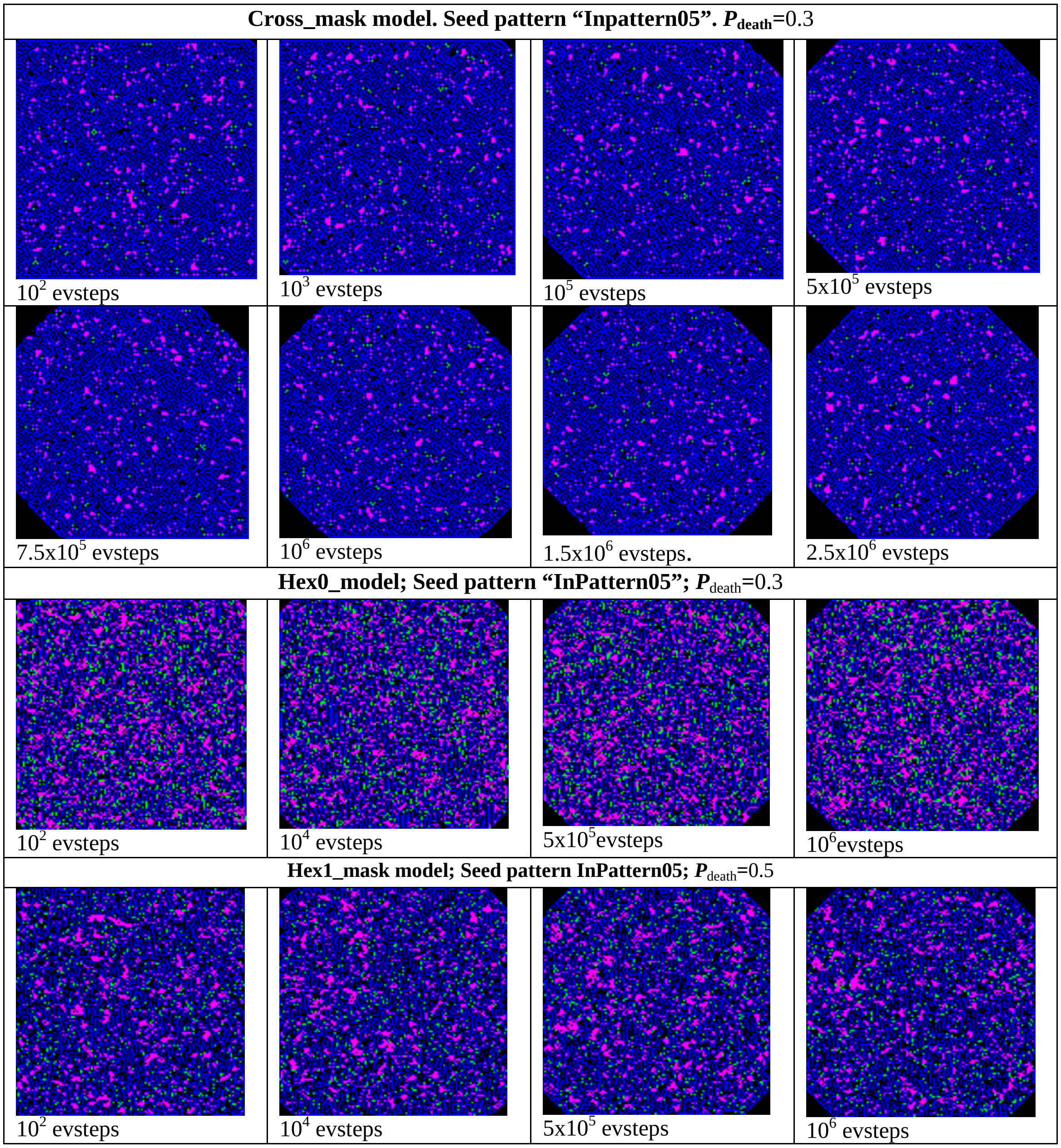

Figure 28. Cross, Hex0 and Hex1_mask models: “eternal life” in a bounded space for various probabilities of “death” and “chaotic” seed patterns. Cells that will “die” on the next step are shown pink, cells that will give “birth” are shown green and stable cells are shown blue; “empty” cells are shown black.

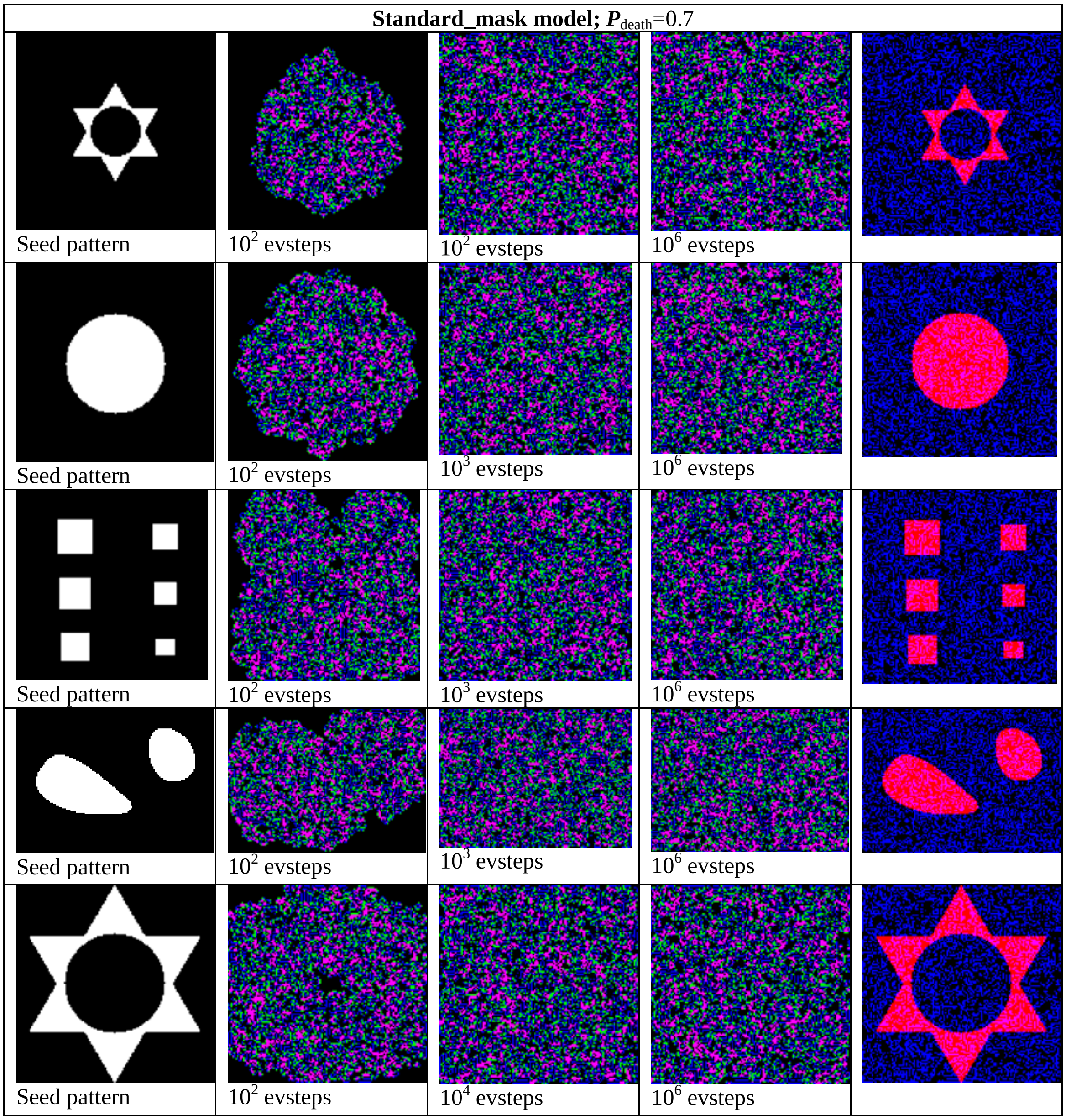

Figure 29. Standard_mask model: "eternal life" in the "vital space". In the last images of every row, shown blue are emerged after $10^6$ evsteps patterns and shown red are corresponding seed patterns. In color coded images in three middle columns cells that will die on the next step are shown pink, cell that will give a birth are shown green and stable cells are shown blue, "empty" cells are shown black; in black and white images "live" cells are shown white.

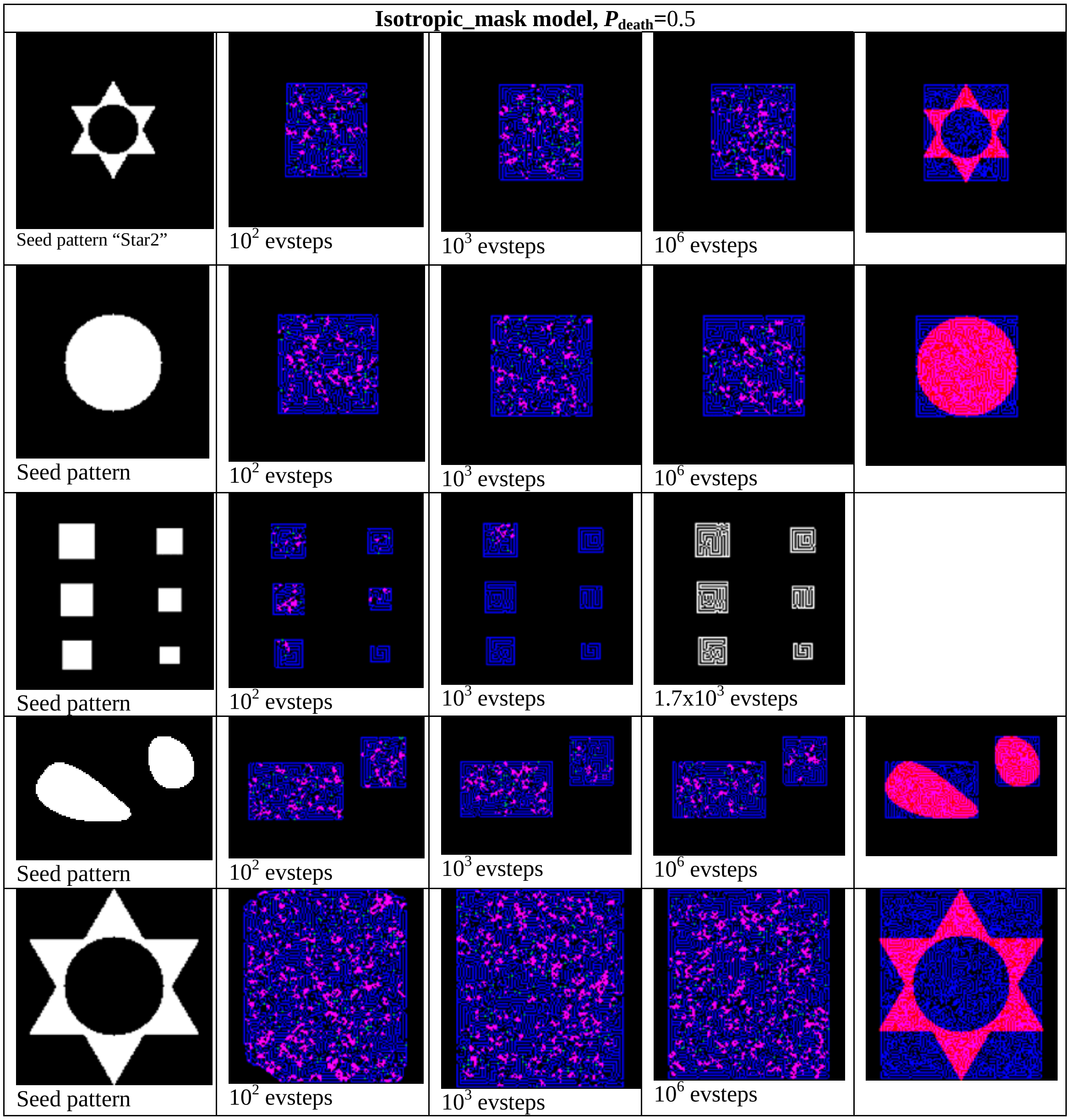

Figure 30. Isotropic_mask model: "eternal life in a bounded space" for solid seed patterns. Last images in each row demonstrate that outer bounds of emerged patterns (shown blue) circumscribe corresponding seed patterns. In color coded images in three middle columns cells that will die on the next step are shown pink, cell that will give a birth are shown green, stable cells are shown blue and "empty" cells are shown black. In color coded images of the right hand column seed patterns are shown red; and emerged patterns after $10^6$ evsteps are shown blue. In black and white images "live" cells are shown white.

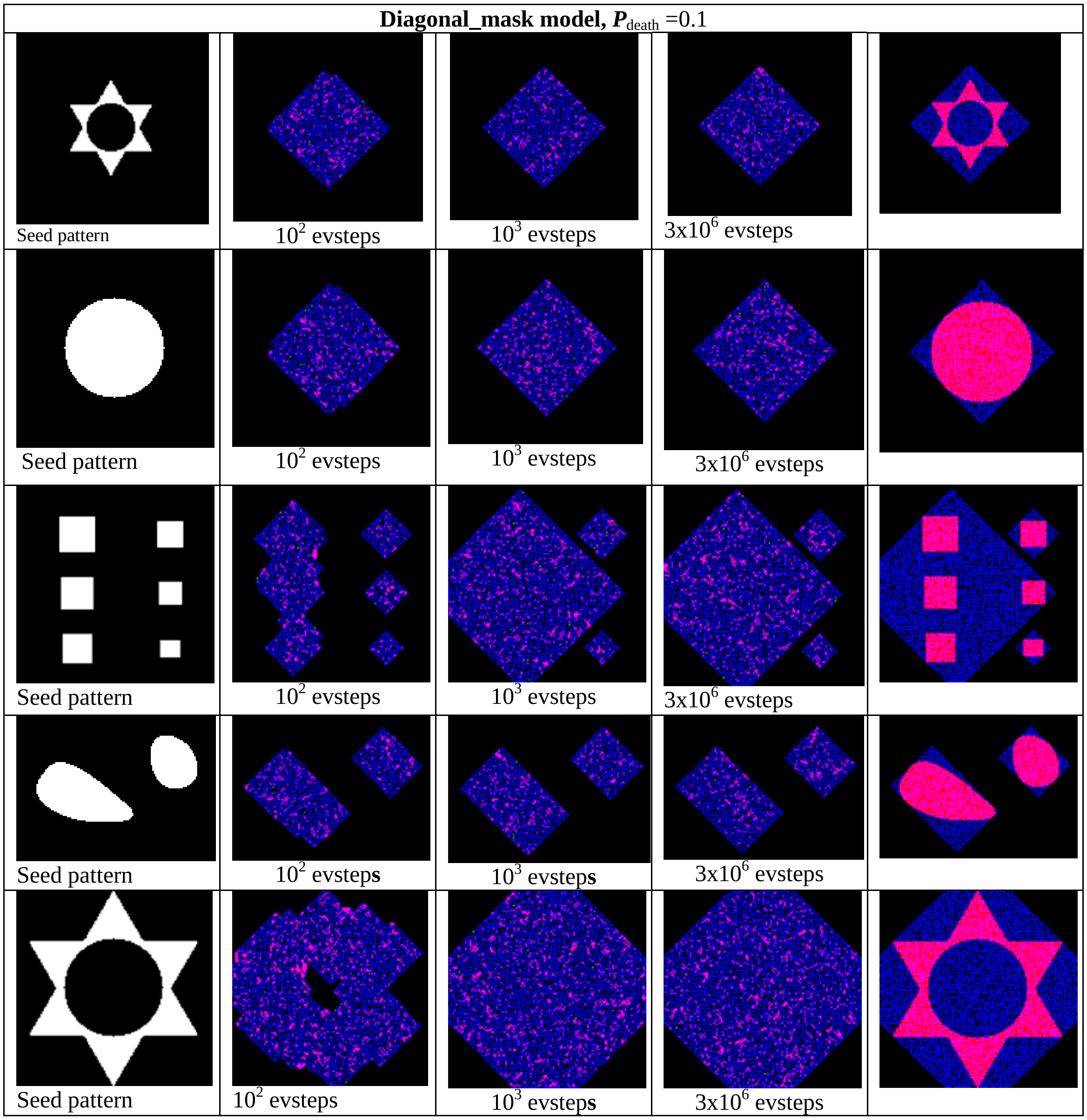

Figure 31. Diagonal_mask model: "eternal life" in a bounded space for solid seed patterns. Last images in each row demonstrate that outer bounds of emerged patterns (shown blue) circumscribe corresponding seed patterns. In color coded images in three middle columns cells that will die on the next step are shown pink, cell that will give a birth are shown green, stable cells are shown blue and "empty" cells are shown black. In color coded images of the right hand column seed patterns are shown red; and emerged patterns after $10^6$ evsteps are shown blue. In black and white images "live" cells are shown white.

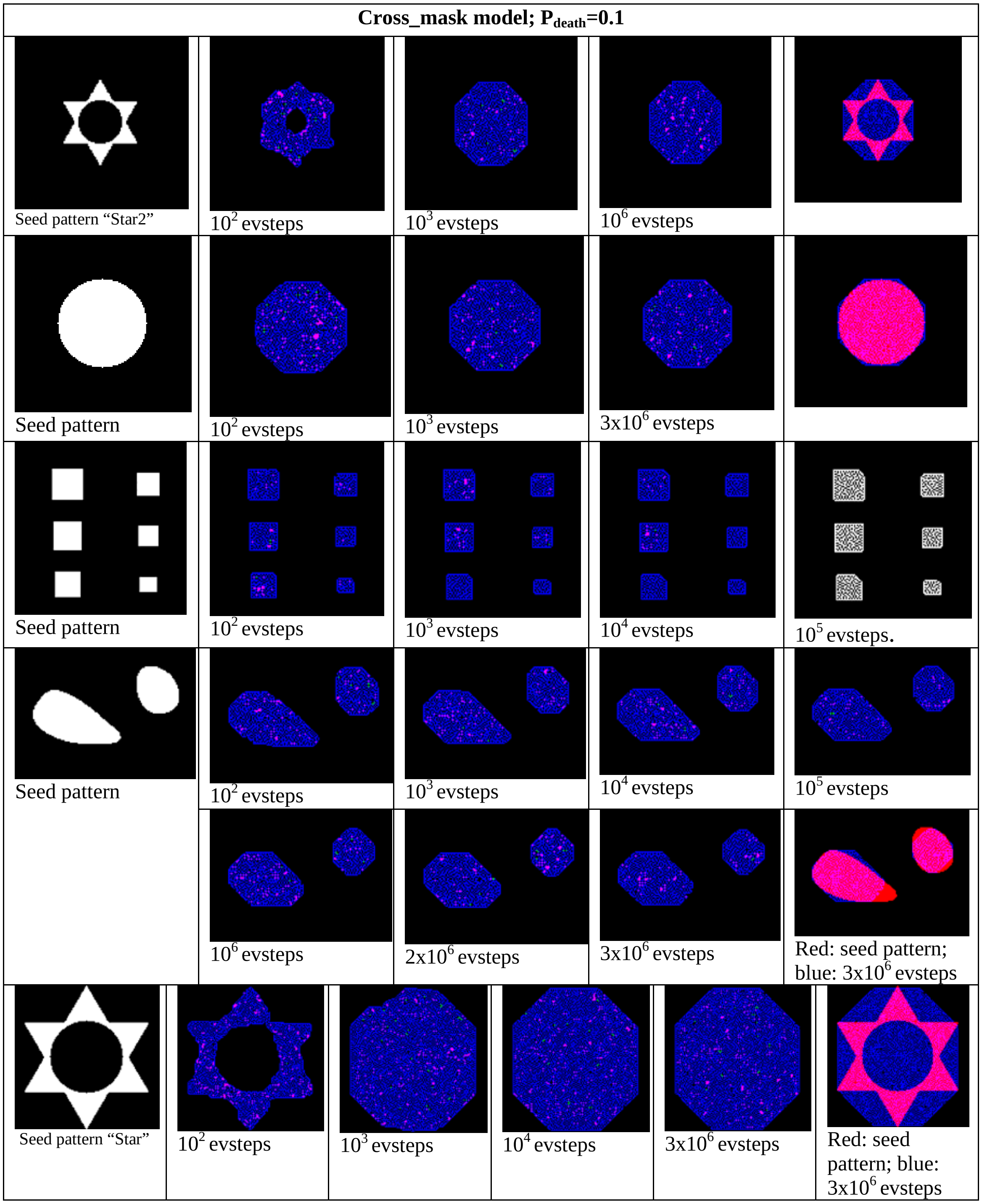

Figure 32. Cross_mask model: "eternal life" in a bounded space is possible when bounded "communities" have sufficiently large size (compare first two rows with last three rows). In color coded images except last images of each sequence cells that will die on the next step are shown pink, cell that will give a birth are shown green, stable cells are shown blue. "Empty" cells are everywhere shown black. In black and white images "live cells are shown white. One can see that for all seed patterns but "Blobs" one, outer bounds of emerged patterns circumscribe corresponding seed patter. This is not the case for seed pattern "Blobs", perhaps due to insufficient size of the blobs.

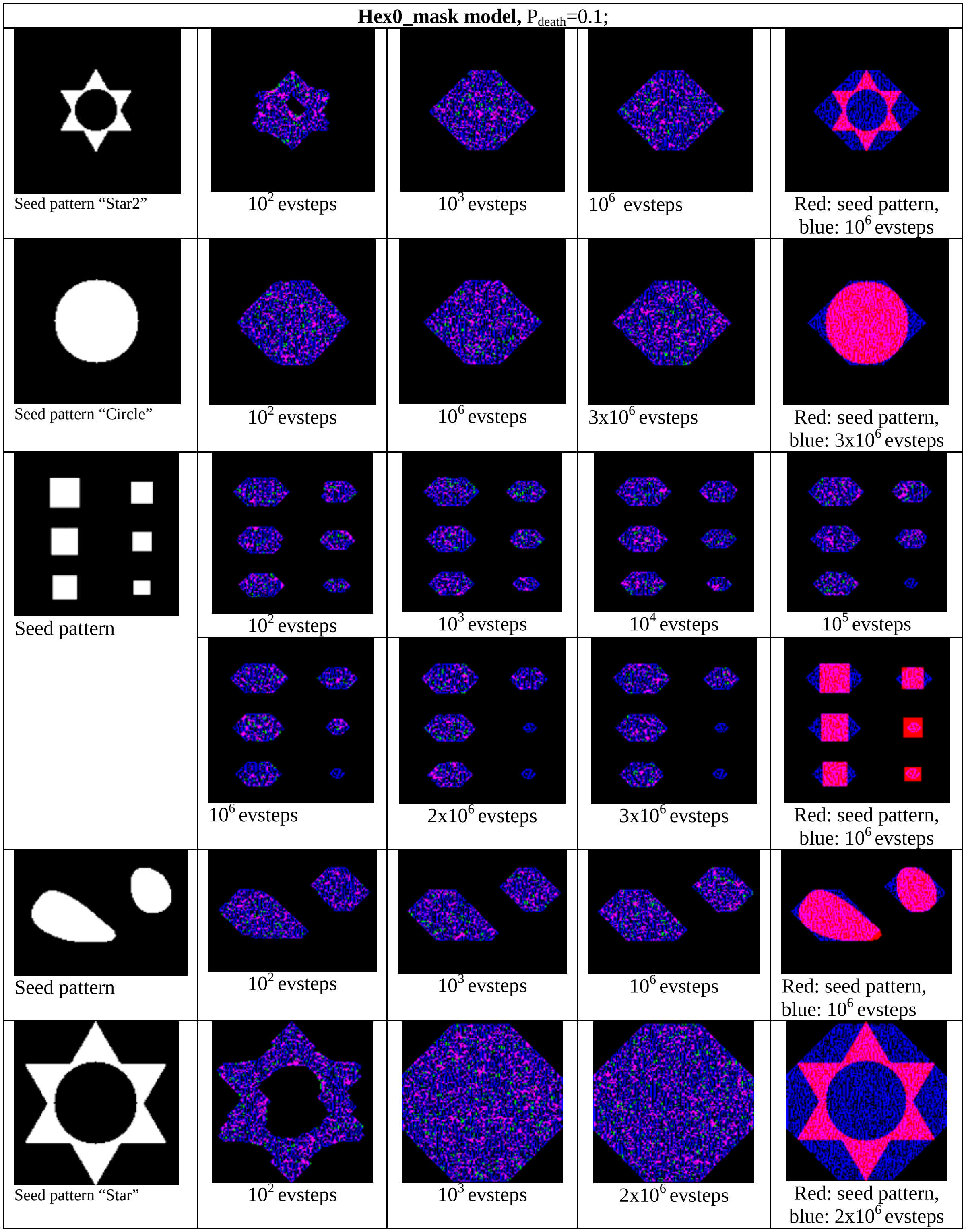

Figure 33. Hex0_mask model: “eternal life” in a bounded space for solid seed patterns. In color coded images except last images in each sequence cells that will die on the next step are shown pink, cell that will give a birth are shown green and stable cells are shown blue. In all images “empty” cells are shown black. For sufficiently large seed patterns, outer bounds of emerged patterns circumscribe the corresponding seed pattern.

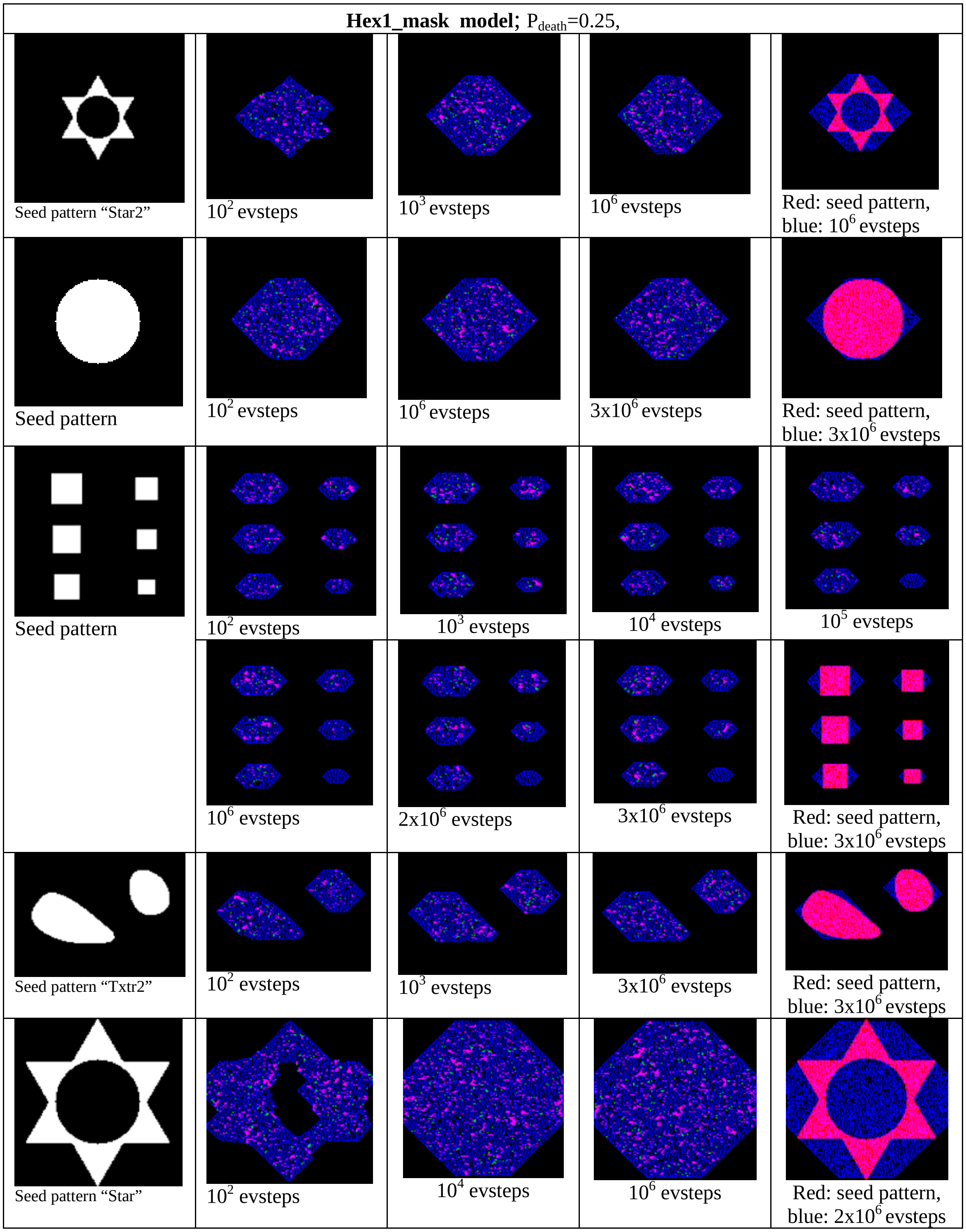

Figure 34. Hex1_mask model: “eternal life” in a bounded space for solid seed patterns. In color coded images except last images in each sequence cells that will die on the next step are shown pink, cell that will give a birth are shown green and stable cells are shown blue. In all images “empty” cells are shown black. For sufficiently large seed patterns, outer bounds of emerged patterns circumscribe the corresponding seed pattern.

One can see from the figures that whereas for Standard_mask and Isotropic_mask models formations growing from seed pattern propagate till they reach bounds of the “vital space” and then stay active in this space, which demonstrates a capability of “unlimited” expansion, dynamics of other models is different. For Diagonal_mask, Cross_mask, Hex0_mask, Hex1_mask and Hex2_mask models, growing formation reach shapes bounded by geometric figures characteristic for each particular model: rectangle or right angles (Isotropic_mask and Diagonal_mask models), octagon (Cross_mask model), hexagon (Hex0_mask ans Hex1_mask models). Remarkably, formation, from chaotic and sufficiently dense seed patterns, of stable in size and shape active “communities” starts from shrinkage of the “vital space” from corners and the shrinkage stops when the “community” reaches a certain “critical” size (see Figure 27, two upper rows, and Figure 28). It is also remarkable that, as a rule, outer bounds of formations emerged from “solid” and sufficiently large seed patterns circumscribe corresponding seed patterns.

One can also see in

Figure **30**, third row,

Figure **30**, third row, Figure 33, third and fourth rows,

Figure **34**, third and fourth rows that the “eternal life in a bounded space” dynamics is possible only if bounded formations reach a sufficiently large size.

Surprisingly, Hex2_mask model seems to be, as Figure 35 shows, incapable of generating patterns that “live” permanently. For very low probabilities of death, its evolution does not enter into the “coherent shrinkage” stage and ends up with maze-like stable patterns even faster than for larger probabilities of “death”.

In most of the experiments, when “eternal life in a bounded space“ dynamics seemed to be observed, we ran models maximum (1÷3)x$10^6$ evolution steps. In order to be better convinced in the possibility of the “eternal life” dynamics, we ran the cross_mask model 2x$10^7$ evolution steps for $\boldsymbol{P}_{death}$=0.3425. The results illustrated in Figure 36 tell in favor of this possibility.

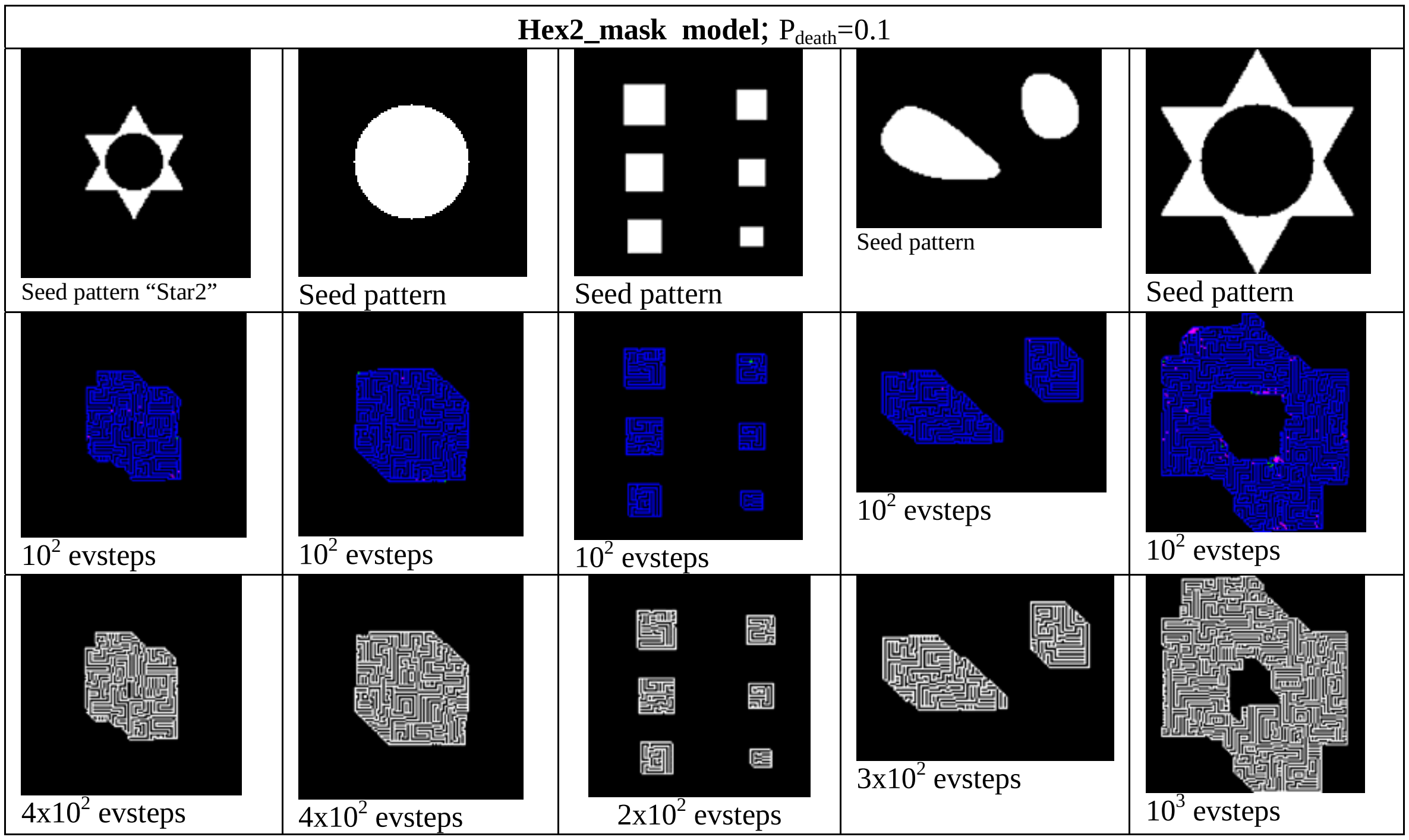

Figure 35. Evolution of the Hex2_mask model for the probability of death 0.1. In color coded images cells that will “die” on the next step are shown pink, cells that will give a birth are shown green and stable cells are shown blue. In black and white images “live” cells are shown white. In all images “empty” cells are shown black.

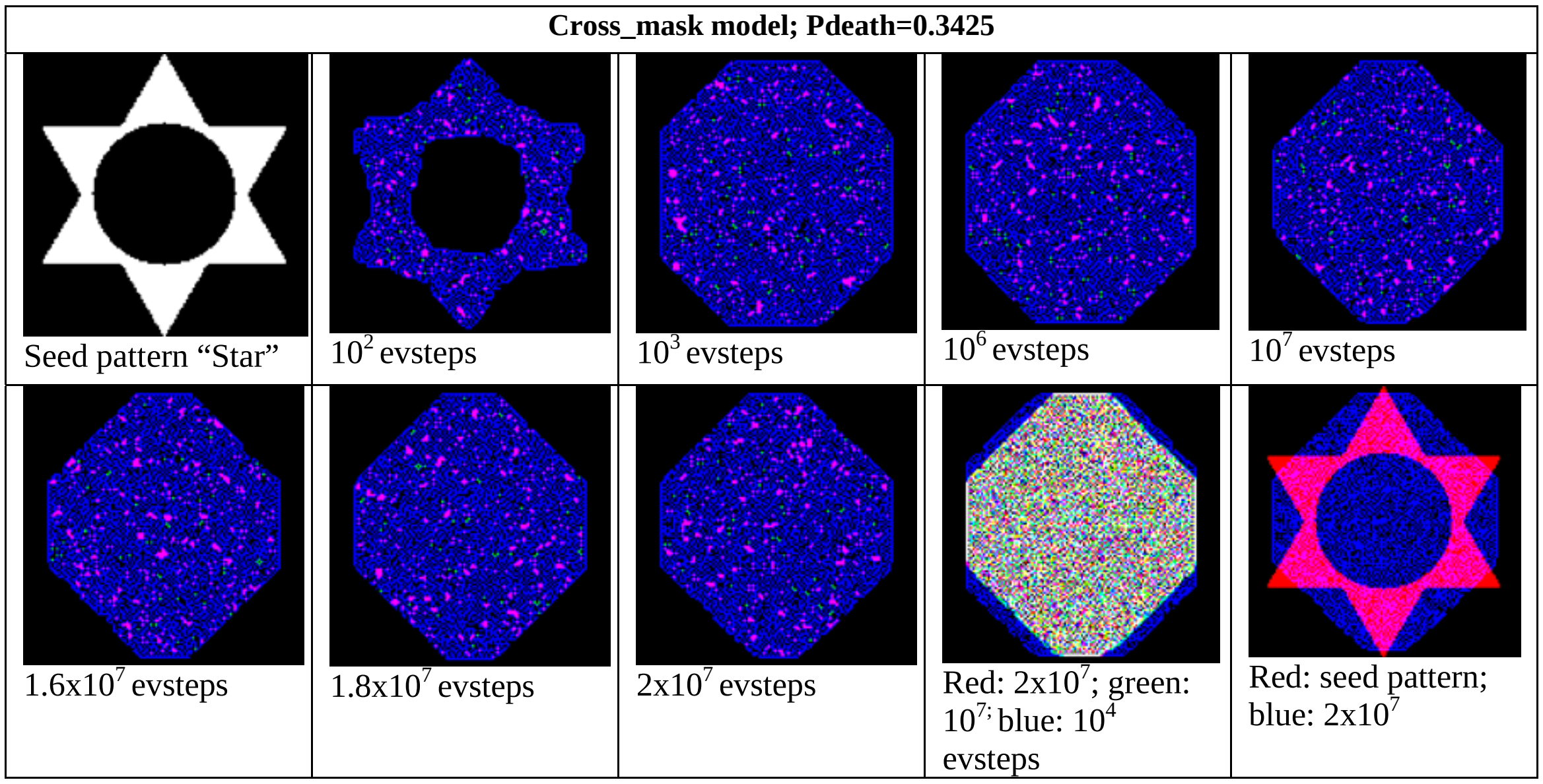

Figure 36**.** Verification of the possibility of “eternal life” in a bounded space dynamics on an example of extra-long evolution of the Cross_mask model**.** Fourth and fifth images in the second row are color coded combinations of patterns emerged after different number of evolution steps in order to demonstrate that patterns do shrink between $10^4$ evsteps and $10^7$ evsteps and practically do not shrink between $10^4$ evsteps and $2x10^7$ evsteps. The very last image in the second row shows, in red, the seed pattern and, in blue, the pattern after $2x10^7$ evsteps and demonstrates that the outer shape of the pattern does not circumscribes the seed pattern as it was observed in the case of $\boldsymbol{P}_{death}$=0.1 (see

Figure 32, lower row).

## *Conclusion*

Several modifications of the standard Conway's Game of Life have been suggested and evolutionary dynamics of the introduced new models has been experimentally investigated. In the experiments, a number of new phenomena has been revealed. Specifically it has been found that

- Standard Conway's model in its stochastic modification with the probability of "death" lower than one demonstrates two types of the evolutionary dynamics:
  - (i) "eternal life" in the "vital space" in the range of the probabilities of "death" $0.3 <\approx P_{death} < 1 - \varepsilon$ , where $\varepsilon$ - is an arbitrarily small number.
  - (ii) "ordering of chaos" into maze-like patterns with stochastic "dislocations" for lower probabilities of death $P_{death} <\approx 0.3$. A remarkable feature of these patterns is that, being stable in the "vital space", their fragments preserve a capability of growth and implantation into other maze-like patterns. These patterns remind patterns of magnetic domains, finger prints, zebra skin, tiger fur, fish skin patterning and alike, which can be frequently found in live as well as in inanimate nature.
- Other introduced models with modified weights exhibit four types of evolutionary dynamics:
  - (iii) For $P_{death} = 1$, they, similarly to the standard non-stochastic Conway's model, exhibit "ordering of chaos" into certain stable or oscillating formation, specific for each model.
  - (iv) For sufficiently large $P_{death} < 1$, the models feature "ordering of chaos" into maze-like patterns or "Manhattan"-like patterns; however, unlike the standard Conway's model, fragments of these patterns extracted from "mother" patterns, have only limited potentials of growth.
  - (v) For intermediate values of $P_{death} < 1$, the models exhibit "eternal life in a bounded space" type of dynamics within "communities" bounded by shapes specific for each model (by squares or right angles, octagons, hexagons oriented parallel to the model rectangular lattice axes or $45^o$ rotated with respect to the lattice axes) and reached through a process of "self-controlled growth"; bounded formations demonstrate seemingly permanent activity (cell births and deaths) while keeping their outer bounds.

(vi) For sufficiently low probabilities of “death”, the models feature dynamics that consist of three stages: (1) “self-controlled growth” into active (in terms of births and deaths) “communities” bounded by shapes, characteristic for each model, (2) “stabilized in shape state of maturity” and (3) “coherent shrinkage” when bounded formations gradually shrink to nil or to a few stable or oscillating formations keeping in this process isomorphism of their bounding shapes until the very end.

Inter alia, these results are remarkable illustrations of how purely local connection between interactive cells can determine their collective behavior.